\documentclass[twocolumn,twocolappendix]{aastex63}

\usepackage{amsmath, amsthm}
\usepackage{amsfonts,amssymb}
\usepackage{cancel}

\submitjournal{ApJL}

\shorttitle{Turbulent viscosity acting on the equilibrium tidal flow in convective stars}
\shortauthors{J. Vidal and A. J. Barker}

\begin{document}

\title{Turbulent viscosity acting on the equilibrium tidal flow in convective stars}

\correspondingauthor{J\'er\'emie Vidal}
\email{j.n.vidal@leeds.ac.uk}

\author[0000-0002-3654-6633]{J\'er\'emie Vidal}
\affiliation{Department of Applied Mathematics, University of Leeds, Leeds, LS2 9JT, UK}

\author[0000-0003-4397-7332]{Adrian J. Barker}
\affiliation{Department of Applied Mathematics, University of Leeds, Leeds, LS2 9JT, UK}




\begin{abstract}
Convection is thought to act as a turbulent viscosity in damping tidal flows and in driving spin and orbital evolution in close convective binary systems. 
This turbulent viscosity should be reduced, compared to mixing-length predictions, when the forcing (tidal) frequency $|\omega_t|$ exceeds the turnover frequency $\omega_{cv}$ of the dominant convective eddies. 
However, two contradictory scaling laws have been proposed and this issue remains highly disputed. 
To revisit this controversy, we conduct the first direct numerical simulations (DNS) of convection interacting with the equilibrium tidal flow in an idealized global model of a low-mass star. 
We present direct computations of the turbulent effective viscosity, $\nu_E$, acting on the equilibrium tidal flow.
We unexpectedly report the coexistence of the two disputed scaling laws, which reconciles previous theoretical (and numerical) findings.
We recover the universal quadratic scaling 
$\nu_E \propto (|\omega_t|/ \omega_{cv})^{-2}$ in the high-frequency regime $|\omega_t|/ \omega_{cv} \gg 1$. 
Our results also support the linear scaling $\nu_E \propto (|\omega_t|/ \omega_{cv})^{-1}$ in an intermediate regime with $1 \leq |\omega_t| / \omega_{cv} \lesssim \mathcal{O}(10)$. 
Both regimes may be relevant to explain the observed properties of close binaries, including spin synchronization of solar-type stars and the circularization of low-mass stars. 
The robustness of these two regimes of tidal dissipation, and the transition between them, should be explored further in more realistic models.
A better understanding of the interaction between convection and tidal flows is indeed essential to correctly interpret observations of close binary stars and short-period planetary orbits. 
\end{abstract}


\section{Introduction}
\label{sec:intro}
Tidal interactions determine the orbital and spin evolution of short-period planets and low-mass binary stars \citep[e.g.][]{Mazeh2008,zahn2008tidal,ogilvie2014tidal}. 
A major weakness of tidal theory is in modeling how tidal flows interact with convection.
Turbulent convection is believed to act as an effective turbulent viscosity in damping large-scale tidal flows \citep[e.g.][]{zahn1966marees}. 
This mechanism is usually invoked to explain the circularization and synchronization of binary systems containing low-mass or solar-like main-sequence stars  \citep[e.g.][]{zahn1989tidal,meibom2005,meibom2006observational,van2016orbital,lurie2017tidal,triaud2017eblm,von2019eblm}, and evolved stars 
\citep[e.g.][]{verbunt1995tidal,Beck2018,price2018binary}. 

The turbulent viscosity, $\nu_E$, is usually estimated by neglecting the oscillatory nature of the tidal flow. 
This leads to $\nu_E \simeq \nu_{cv}$, where $\nu_{cv}$ is predicted by mixing-length theory \citep[MLT; e.g.][]{Spiegel1971}.
However, when the tidal frequency $|\omega_t|$ is faster than the turnover frequency $\omega_{cv}$ of the dominant convective eddies, $\nu_{E}$ ought to be reduced, as recognized initially by \citet[][]{zahn1966marees}. 
The magnitude of this reduction remains highly disputed.
Two scaling laws that are based on phenomenological arguments have been proposed, with either a linear reduction $\nu_E \propto \nu_{cv} \, (|\omega_t| / \omega_{cv})^{-1}$ \citep{zahn1966marees,zahn1989tidal}, or a quadratic suppression $\nu_E \propto \nu_{cv} \, (|\omega_t| / \omega_{cv})^{-2}$ \citep{goldreich1977solar,goldreich1977turbulent}. 
Revisiting this controversy has been attempted recently by using direct numerical simulations (DNS). 
The two laws have only been recovered in separate studies, which support either the linear scaling \citep[][]{penev2007dissipation,penev2009dissipation,penev2009direct}
or the quadratic one \citep[][]{ogilvie2012interaction,braviner2016stellar,duguid2019tides}. 
Thus, any application of tidal theory to stars (or planets) with convection zones remains uncertain.
Resolving this issue is essential before we can apply tidal theory to interpret observations of close binaries \citep[e.g.][]{kirk2016kepler,lurie2017tidal,van2016orbital,triaud2017eblm,price2018binary} or short-period planetary orbits \citep[e.g.][]{Rasio1996}.
For instance, circularization of subgiant stars with orbital periods of approximately one day could occur in either $\sim 10^2$ or $10^6$ yr, depending on which scaling is valid \citep{price2018binary}. 

Owing to the importance of this problem to interpret observations, we revisit this controversy using global numerical simulations. 
So far, only numerical studies using local models and with simplified imposed shear flows have been undertaken. 
Local DNSs may not capture the full complexity of the tidal response existing in a global model.
They also could be affected by the adopted boundary conditions. 
We therefore set out to gain independent physical insight from global DNSs of convection in the presence of more realistic tidal flows.
This Letter is organized as follows. 
We present our global model in Section \ref{sec:model}. 
Direct computations of the turbulent viscosity are presented in Section \ref{sec:results}. 
The implications and astrophysical extrapolation of our results are presented in Section \ref{sec:discussion}, and we conclude the Letter in Section \ref{sec:ccl}.

\section{Description of the tidal problem}
\label{sec:model}
We study the interplay between tides and turbulent convection in a global model of a low-mass star (or core-less giant planet). 
The primary body is a full sphere of radius $R$, filled with a fluid of uniform (laminar) kinematic viscosity $\nu$ and thermal diffusivity $\kappa$. 
This body is subjected to tidal forcing from an orbiting companion. 
We model convection in the Boussinesq approximation, studying slight departures from a motionless conduction state sustained by homogeneous internal heating. 
Since many low-mass stars are slow rotators \citep[e.g.][]{nielsen2013rotation,Newton2018}, 
and for simplicity, we neglect rotation in this study. 
We define the temperature perturbation $\Theta$ and the velocity field $\boldsymbol{u}+\boldsymbol{U}_0$, where we split the flow into a background large-scale tidal flow $\boldsymbol{U}_0$ and a perturbation $\boldsymbol{u}$.
We use dimensionless units for the simulations, adopting $R$ as our length scale and ${R^2}/{\nu}$ as our timescale.
Convection is then governed by two dimensionless numbers, the Rayleigh number $Ra$ (which measures the strength of the convective driving) and the Prandtl number $Pr=\nu/\kappa$. 
We solve the system of equations in their weak variational form by using the spectral-element code Nek5000 \citep{fischer2007simulation}, employed previously for tidal studies \citep[e.g.][]{favier2014non,barker2016nonb,reddy2018turbulent}.  
Further details of the model are given in Appendix \ref{appendix:convection}. 

\begin{figure}
    \centering
    \begin{tabular}{c}
    \includegraphics[width=0.4\textwidth]{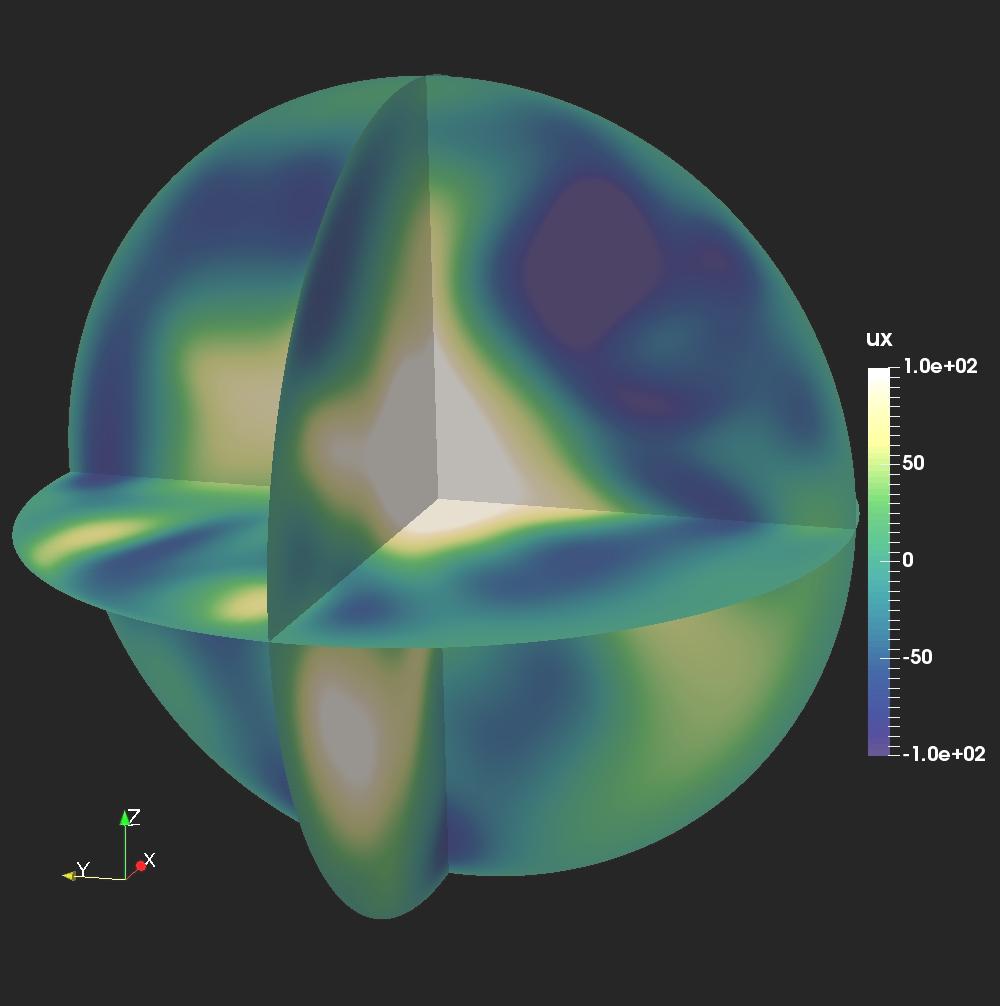} \\
    \includegraphics[width=0.49\textwidth]{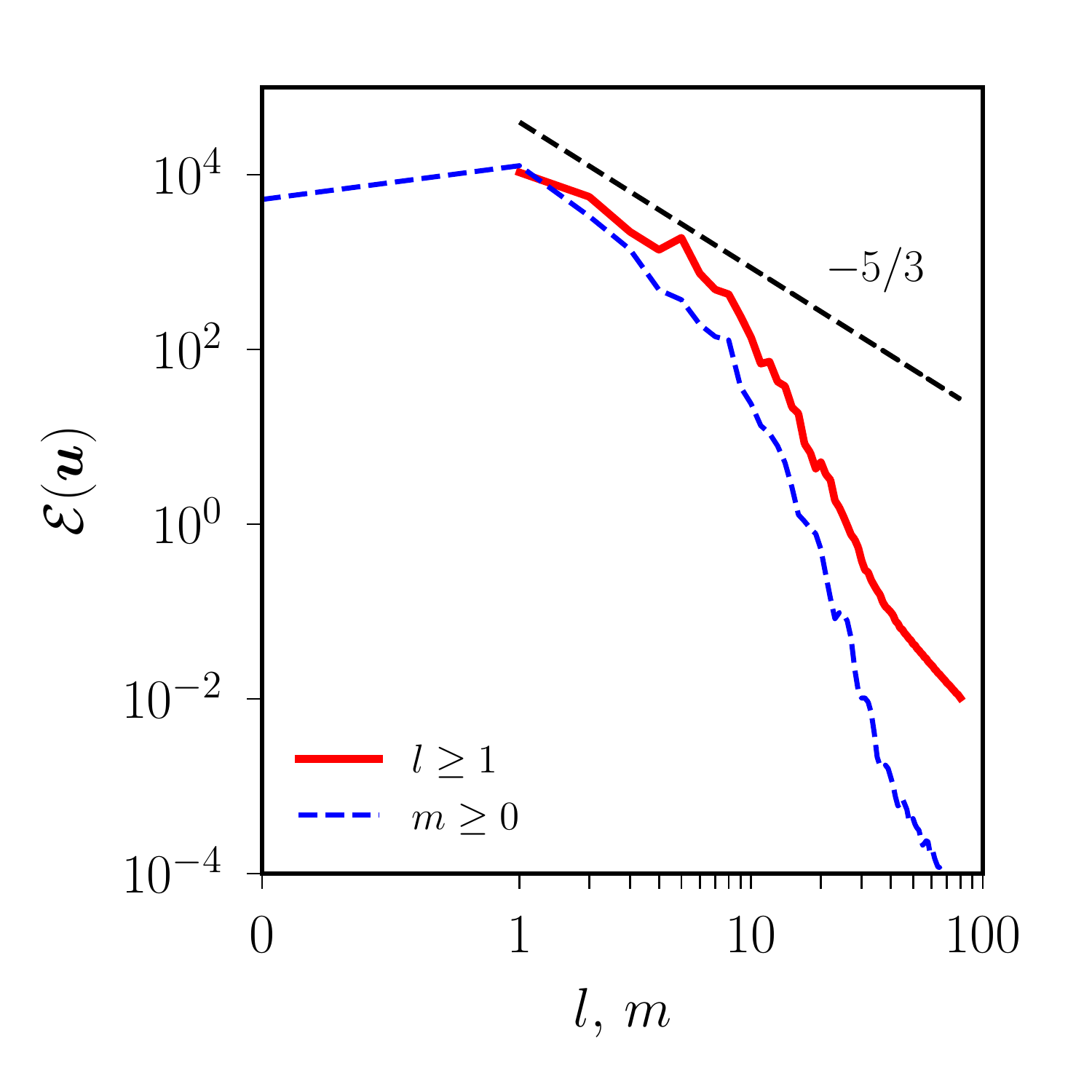} \\
    \end{tabular}
    \caption{Simulation of uniformly heated turbulent convection without tides ($Ra=10^6, Pr=1$). 
    \emph{Top}: three-dimensional snapshot of the velocity field component $u_x$. 
    \emph{Bottom}: spectra of the instantaneous kinetic energy $\mathcal{E}(\boldsymbol{u})$, as a function of the spherical harmonic degree $l\geq1$ and azimuthal number $m\geq0$ (using orthonormalized harmonics) between radii $r \in [0.05, 0.99]$. 
    Spectra have been computed by interpolating the data to a spherical grid. 
    }
    \label{fig:convection}
\end{figure}

Previous numerical studies modeled the tidal flow with either an (ad hoc) external forcing \citep{penev2009direct} or with a background unidirectional shear flow in a shearing box \citep{ogilvie2012interaction,braviner2016stellar,duguid2019tides}. 
Here, we instead consider self-consistently the large-scale (non-wavelike) equilibrium tidal flow \citep[e.g.][]{remus2012equilibrium}. 
We assume that the companion is a point mass, moving on a circular orbit around the star.  
Thus, the dominant component of the tidal potential in the inertial frame has the spherical harmonic degree $l=2$ and azimuthal order $m=2$ \citep{ogilvie2014tidal}. 
In the inertial frame, the resulting (dimensionless) flow is in the $xy$-plane and takes the form
\begin{equation}
\boldsymbol{U}_0 = -\frac{\omega_t \beta}{2}
	\begin{pmatrix}
	\sin (\omega_t t) & \cos (\omega_t t) \\
    \cos(\omega_t t) & -\sin (\omega_t t) \\
	\end{pmatrix}
	\begin{pmatrix}
	x \\
    y \\
	\end{pmatrix},
    \label{eq:U0}
\end{equation}
where $\beta \ll 1$ is the dimensionless tidal amplitude (roughly the ratio of tidal displacement to unperturbed radius) and $\omega_t$ is the tidal (angular) frequency (twice the orbital frequency in the absence of rotation).
The forcing amplitude $\beta$ must be large enough to obtain a measurable tidal response in the presence of convection, but large values could strongly modify the convection. 

The global simulations that we present here are very demanding, because they must be run for a sufficiently long duration to reduce turbulent noise.
This inevitably restricts our survey of parameter space. 
We simulate highly supercritical convection with $Ra=10^6$ and $Pr=1$, which can be compared with the value for linear onset $Ra\geq 4019$ \citep[computed with a dedicated solver;][]{monville2019rotating}.   
The parameters and outputs for each simulation are given in Appendix~\ref{appendix:convection}. 
The convection in the saturated state without tides (i.e. $\beta=0$) is illustrated in Figure \ref{fig:convection}. 
The kinetic energy is characterized by a nonnegligible axisymmetric component (consistent with the flow in the top panel), and a short inertial-like range illustrated by the Kolmogorov scaling ($\propto -5/3$, bottom panel). 
We quantitatively estimate the convective turnover frequency as $\omega_{cv}=u_{rms}/l_E$, where $u_{rms}$ is the time average of the volume-averaged rms velocity and $l_E\simeq 1/3$ is here the turbulent length scale. 
We obtain a typical value $\omega_{cv} \simeq 143.5 \, \pm \, 3.2$ for the (dimensionless) convective angular frequency, and 
the mean properties of the convection are not significantly affected in the presence of the tidal flow (see Appendix \ref{appendix:results}). 

\section{Turbulent Viscosity}
\label{sec:results}
\begin{figure}
    \centering
    \begin{tabular}{cc}
    \includegraphics[width=0.49\textwidth]{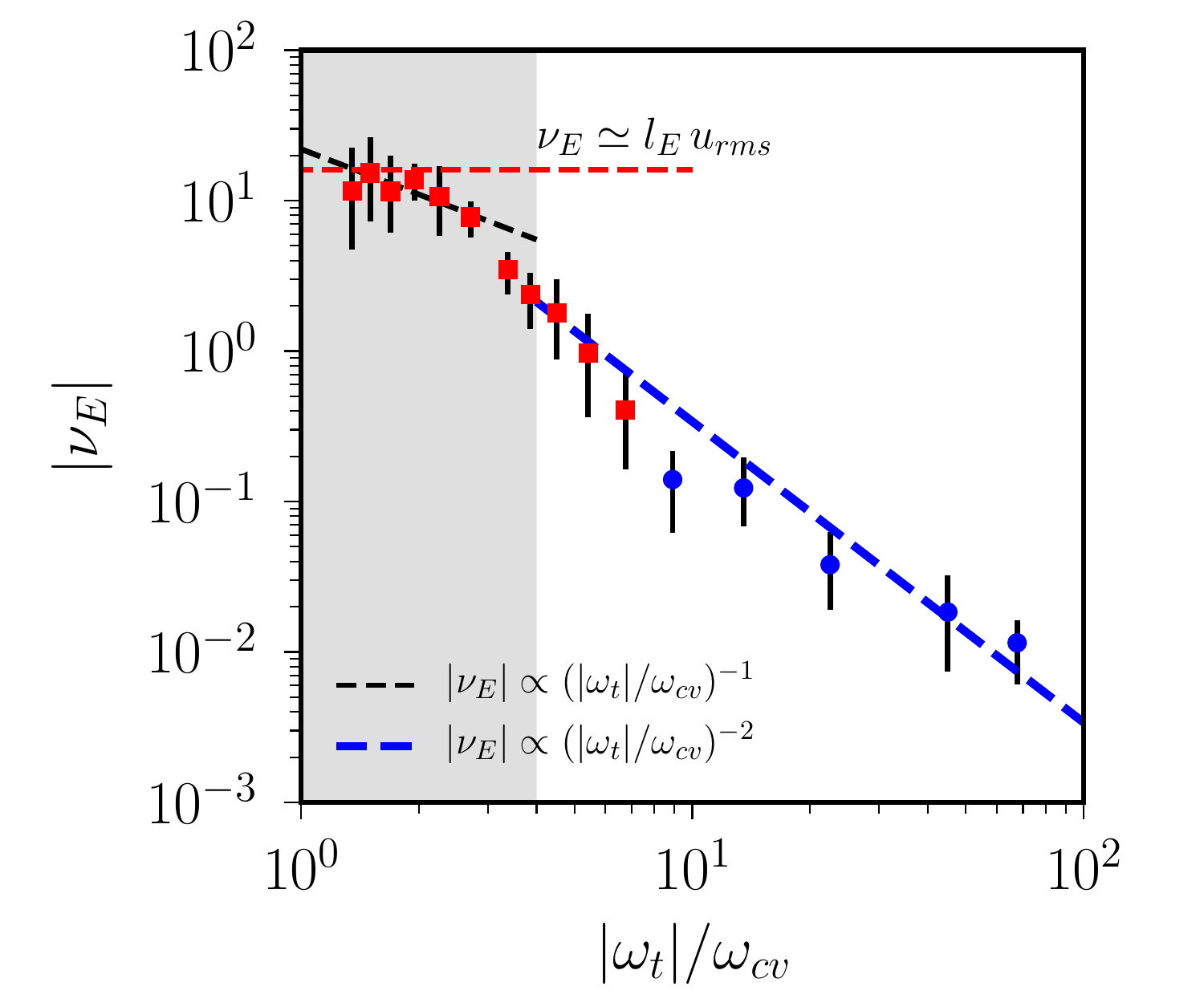} \\
    \includegraphics[width=0.49\textwidth]{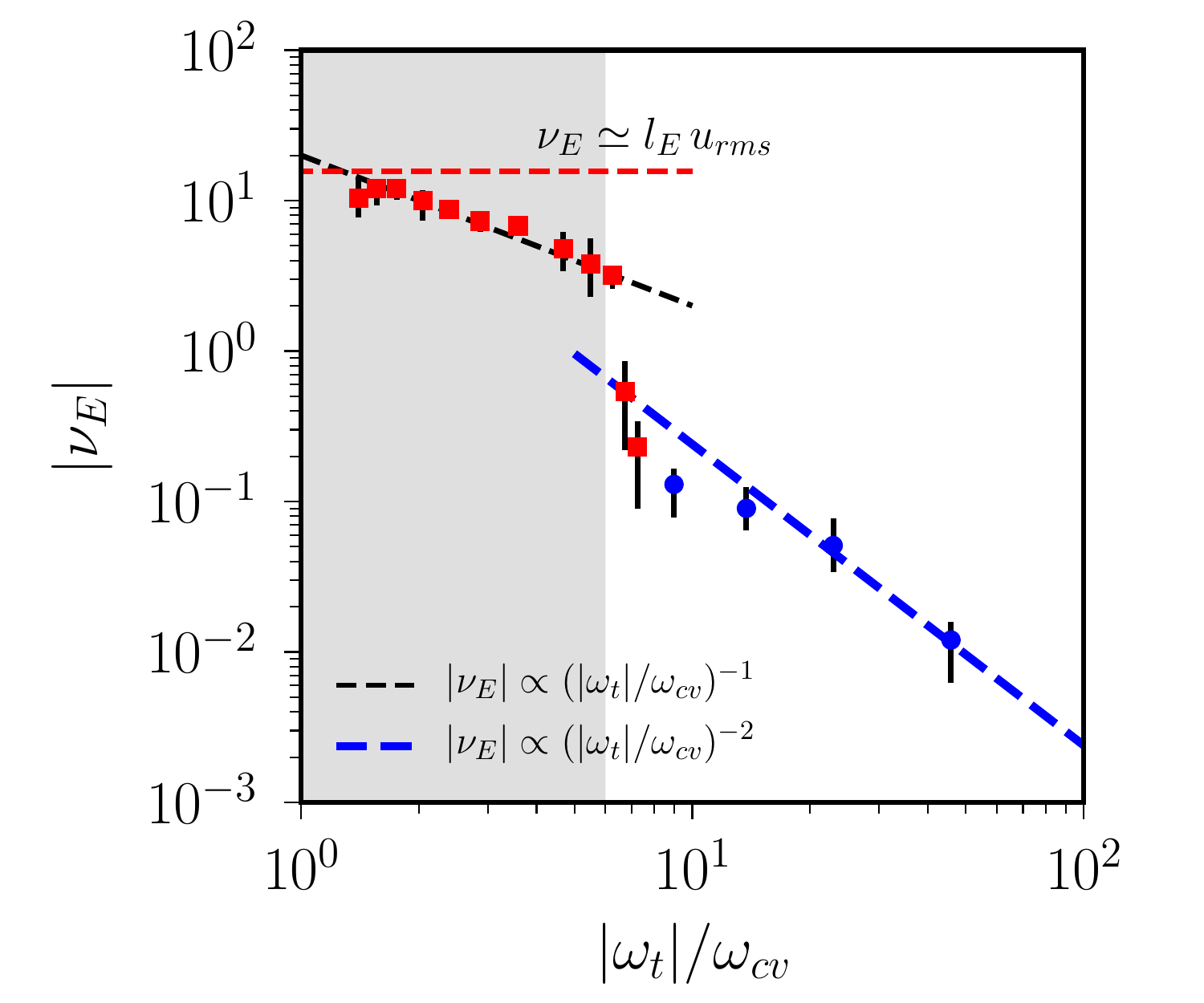} \\
    \end{tabular}
    \caption{Direct measurements of the effective viscosity $\nu_E$ in turbulent convection ($Ra = 10^6 ,Pr=1$), as a function of $|\omega_t|/\omega_{cv}$. Red squares: $\nu_E > 0$. Blue circles: $\nu_E < 0$. Error bars are conservatively defined using two standard deviations from the mean value.     \emph{Top}: $\beta=10^{-2}$.
    \emph{Bottom}: $\beta=5\times10^{-2}$.
    Horizontal dashed lines: expected behavior from MLT $\nu_E \simeq l_E \, u_{rms}$ in the low-frequency regime ($|\omega_t|\ll \omega_{cv}$). 
    }
    \label{fig:eddy}
\end{figure}

We determine numerically the effective viscosity coefficient $\nu_E$, which is the leading-order component of the effective viscosity tensor at the forcing frequency \citep[e.g.][]{penev2009direct}. 
This is obtained numerically by balancing the mean rate at which convection does work on the tidal flow with the mean rate of viscous dissipation of this flow \citep[e.g.][]{goodman1997fast,braviner2016stellar}. 
This leads to
\begin{equation}
   \nu_E =- \frac{1}{(\omega_t \beta)^2 \Delta T} \int_{t_0}^T  \langle \boldsymbol{u} \boldsymbol{\cdot} \left [ (\boldsymbol{u} \boldsymbol{\cdot} \boldsymbol{\nabla}) \, \boldsymbol{U}_0 \right ] \rangle_V \, \mathrm{d} t,
    \label{eq:eddy}
\end{equation}
with $\langle \, \cdot \, \rangle_V = (1/V) \int_V \cdot \ \mathrm{d} V$ the volume average and $5 \leq \Delta T = T-t_0 \leq 10$, with $t_0$ an appropriate initial time in the saturated regime. 
We have verified in Appendix \ref{appendix:results} that the spatial average is not dominated by regions near the boundary, and is instead due to interactions with turbulent flows in the bulk.

Results for the effective viscosity $\nu_E$ are shown in Figure \ref{fig:eddy}, for the tidal amplitudes $\beta=10^{-2}$ and $\beta=5\times 10^{-2}$. 
The former value is similar to that for a solar-mass binary in a one-day orbit. The effective viscosity decreases as we increase  $|\omega_t|/\omega_{cv}$. 
The striking feature here is the coexistence of both heuristic scaling laws.
First, we obtain an intermediate regime $1 \leq |\omega_t|/\omega_{cv} \lesssim \mathcal{O}(10)$, in which $\nu_E$ is consistent with the linear reduction \citep{zahn1966marees,zahn1989tidal}. 
This trend is clearer in the simulations with the largest tidal amplitude ($\beta=5\times10^{-2}$), because the signal-to-noise ratio (S/N) is lower for weaker tides (as shown by the error bars in Figure \ref{fig:eddy}).
Second, we clearly obtain the quadratic law $|\nu_E| \propto (|\omega_t|/\omega_{cv})^{-2}$ in the high-frequency regime $|\omega_t| \gg \omega_{cv}$ \citep{goldreich1977turbulent,goodman1997fast}. 
The transition between these two scalings is sharp, occurring when $|\omega_t|/\omega_{cv} \simeq 6$ for $\beta = 5\times 10^{-2}$ (bottom panel), but appears to depend weakly on the tidal amplitude. 
These results demonstrate that both scaling laws are obtained in our global model, which have only been found previously in separate studies in Cartesian geometry. 

In the low-frequency regime ($|\omega_t| \lesssim \omega_{cv}$), we have been unable to accurately determine $\nu_E$. 
The amplitude of tidal flow (\ref{eq:U0}) was too weak to give a sufficiently strong S/N. 
A crude extrapolation of our results is broadly consistent with MLT, which would predict $\nu_E \propto \nu_{cv} \simeq l_E \, u_{rms}$ when $|\omega_t| \to 0$ (albeit with an uncertain proportionality constant).
This would be consistent with local simulations \citep{duguid2019tides}. 
However, the convective viscosity could be larger than the MLT prediction in that range \citep[e.g.][]{goldman2008effective}. 
 
\section{Discussion}
\label{sec:discussion}
\subsection{Non-Kolmogorov Turbulence?}
\begin{figure*}
    \centering
    \begin{tabular}{cc}
    \includegraphics[width=0.45\textwidth]{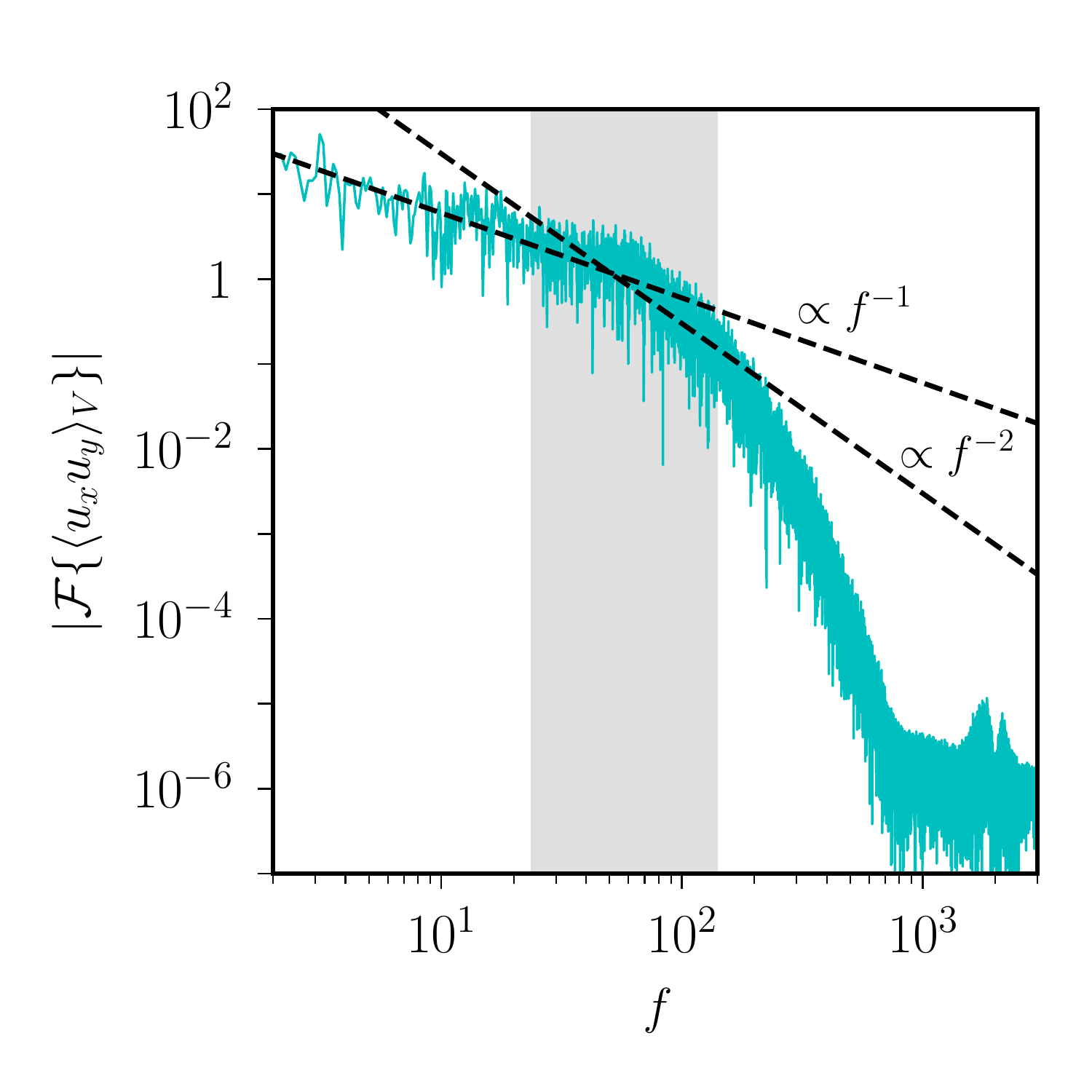} &
    \includegraphics[width=0.45\textwidth]{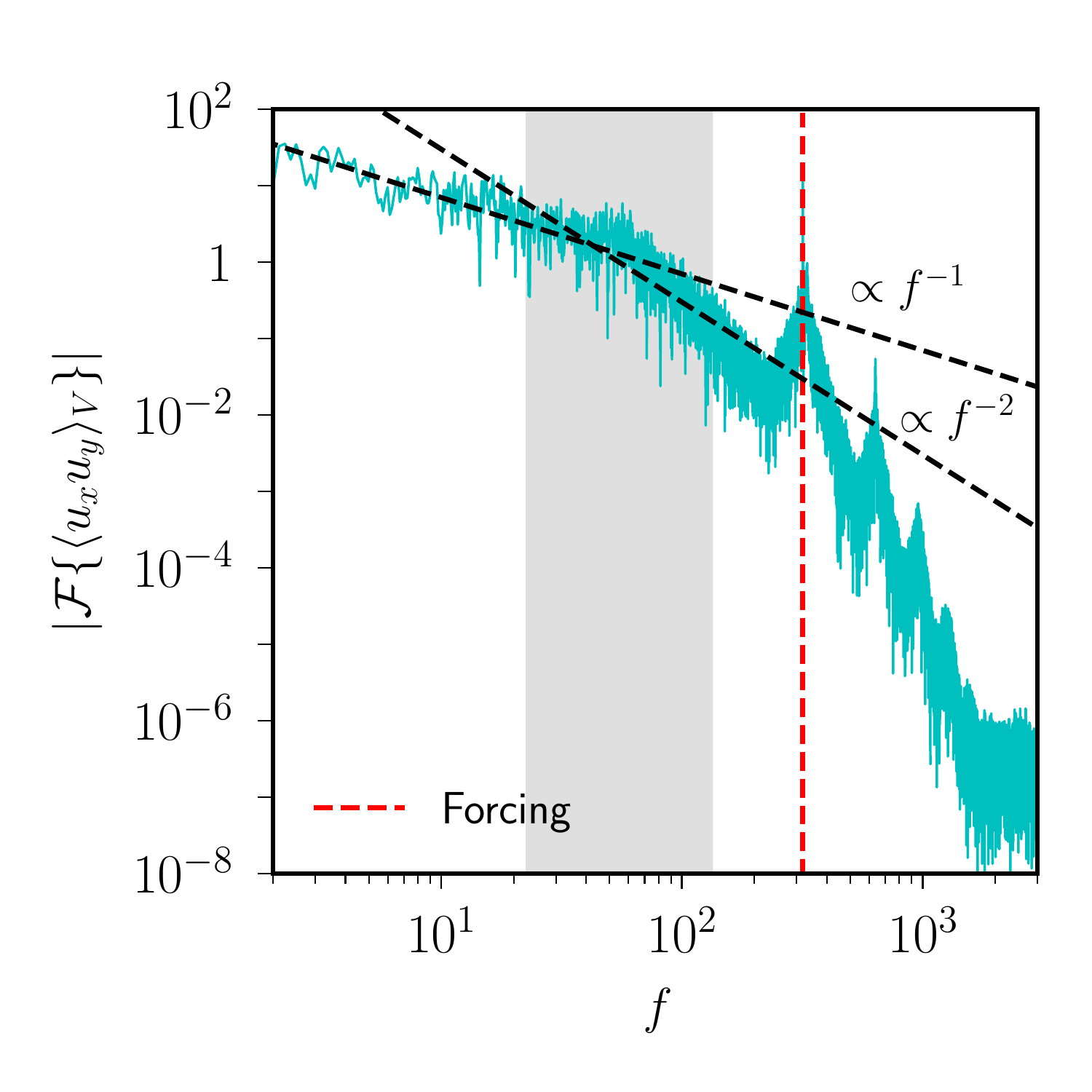} \\
    \includegraphics[width=0.45\textwidth]{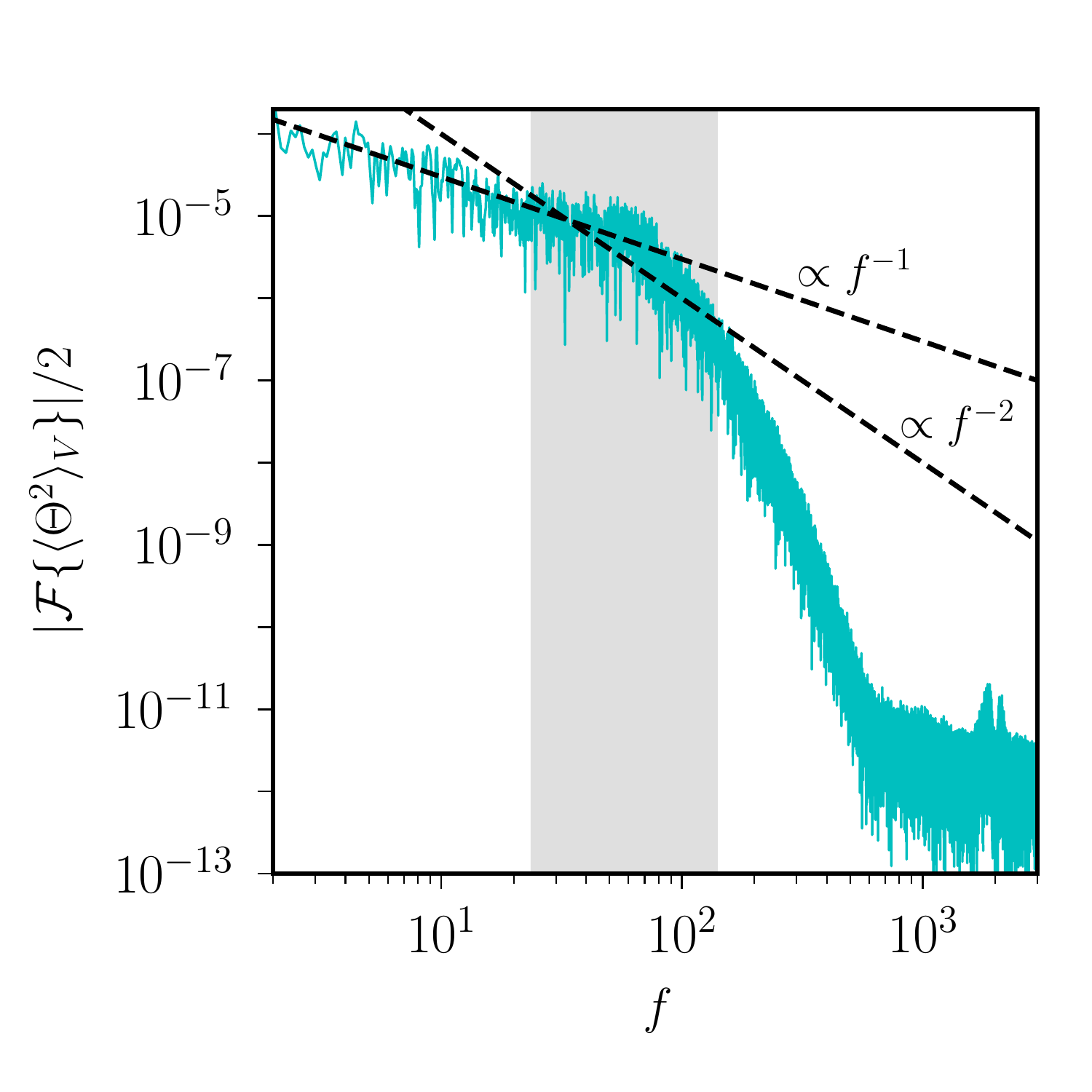} &
    \includegraphics[width=0.45\textwidth]{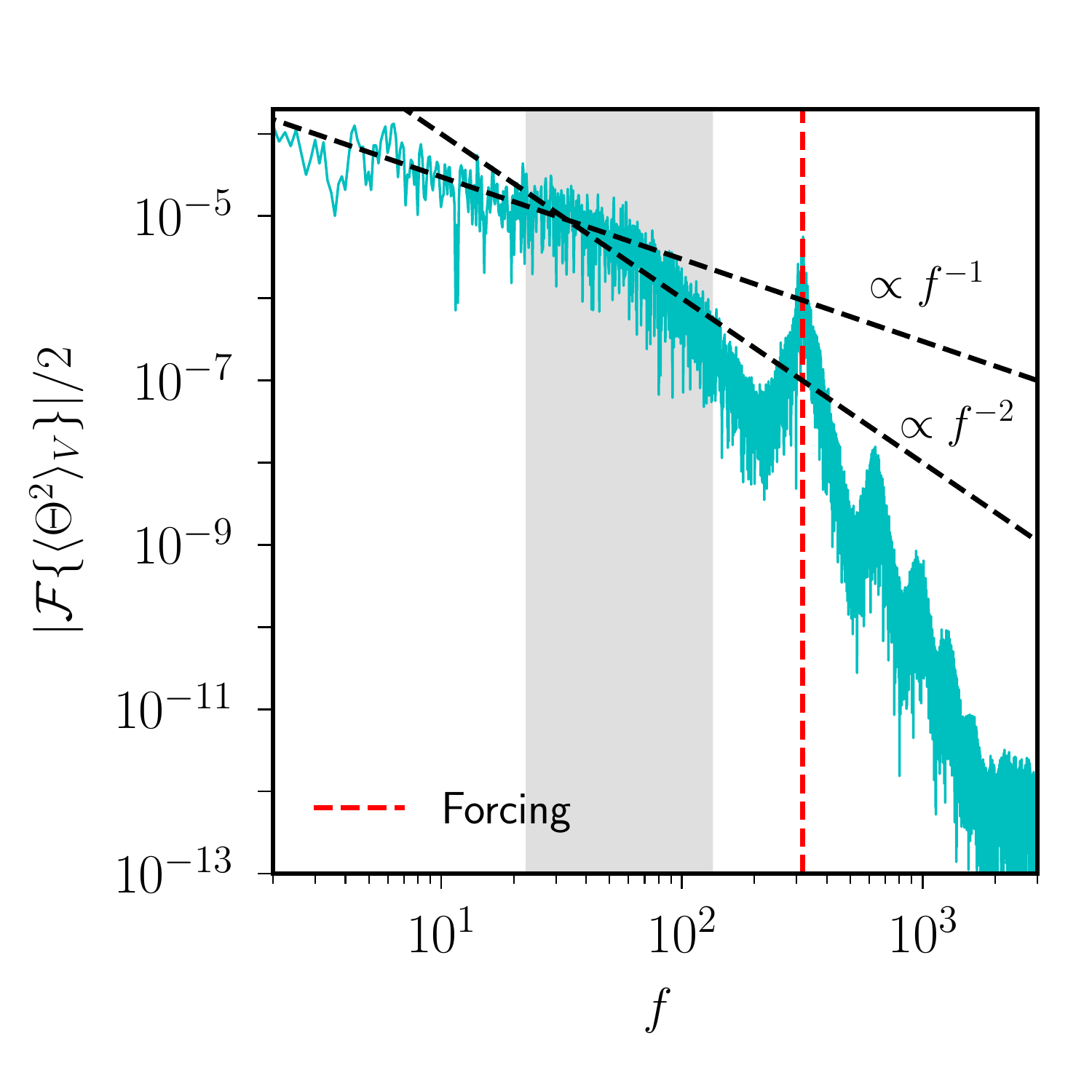} \\
    \end{tabular}
    \caption{Frequency spectra of turbulent 
    convection ($Ra=10^6, Pr=1$). Spectrum of $|\mathcal{F}\{\langle u_x u_y \rangle_V \}|$ (top panel) and $|\mathcal{F}\{\langle \Theta^2 \rangle_V/2 \}|$ (bottom panel), where $\mathcal{F}$ is the Fourier transform and $f$ is the (ordinary) frequency. Gray area shows the intermediate frequency range $1 \leq |\omega_t|/\omega_{cv} \leq 6$ where the linear reduction is observed in Figure \ref{fig:eddy}.
    \emph{Left}: unperturbed convection. 
    \emph{Right}: perturbed convection with $\beta=5 \times 10^{-2}$ and $|\omega_t|/\omega_{cv} = 14.2$.
    }
    \label{fig:spectrum}
\end{figure*}

Turbulent viscosity is often defined with a closure model that relates the Reynolds stress to the rate of strain. 
\citet{ogilvie2012interaction} and \citet{duguid2019tides} demonstrated the viscoelastic nature of the high-frequency ($|\omega_t|/\omega_{cv} \gg 1$) tidal response, developing an asymptotic theory for the Reynolds stress. 
This strongly supports the quadratic reduction for $\nu_E$.
In our global simulations, we have confirmed the viscoelastic character of the response for high-frequency tidal forcing (see Appendix \ref{appendix:results}). 
However, this asymptotic theory does not strictly apply for lower frequencies. 
Indeed, a linear reduction may result from the non-Kolmogorov nature of the turbulence \citep{penev2007dissipation,penev2009dissipation,penev2009direct}. 

We illustrate in Figure \ref{fig:spectrum} the frequency spectrum of the convection. 
The largely non-Kolmogorov nature of the convection is revealed by the frequency spectrum of the volume-averaged Reynolds stress component $\langle u_x u_y\rangle_V$. 
The latter quantity, which is directly related to the effective viscosity (see Appendix \ref{appendix:results}), has a shallower decay with frequency (in $f^{-1}$) than expected from Kolmogorov theory when $|\omega_t|/\omega_{cv} \leq \mathcal{O}(10)$. 
This slope is largely unaffected by the tidal flow, and so is a generic property of the convection in this range. 
The slope of the non-Kolmorogov spectrum is similar to that reported in \citet{penev2007dissipation,penev2009direct}, despite the model differences. 
For larger frequencies, a steeper decay is observed, first behaving like $f^{-2}$ in apparent agreement with local simulations of Rayleigh-Benard convection \citep{kumar2018applicability}, and then rapidly decaying (corresponding with a dissipation range). 
The transition between the linear and quadratic reductions may broadly coincide with where the shallow non-Kolmogorov scaling in $f^{-1}$ ceases to be valid.
Then, even though our spectrum is still non-Kolmogorov-like, a quadratic reduction is found for higher frequencies, in agreement with prior asymptotic theory \citep{ogilvie2012interaction,duguid2019tides}. 
The frequency spectrum of the thermal energy $\langle \Theta^2 \rangle_V/2$ exhibits the same scaling behavior as the Reynolds stress. 
Following \citet{goodman1997fast}, this quantity could also be relevant for the frequency dependence\footnote{As suggested by the referee, based on \citet{phinney1992pulsars}.} of $\nu_E$. 
However, our simulations do not currently allow us to assess their arguments conclusively. 
In summary, our new global simulations support both the linear and quadratic reductions for the eddy viscosity.

\subsection{Astrophysical Implications}
\begin{figure}
    \centering
    \begin{tabular}{c}
    \includegraphics[width=0.49\textwidth]{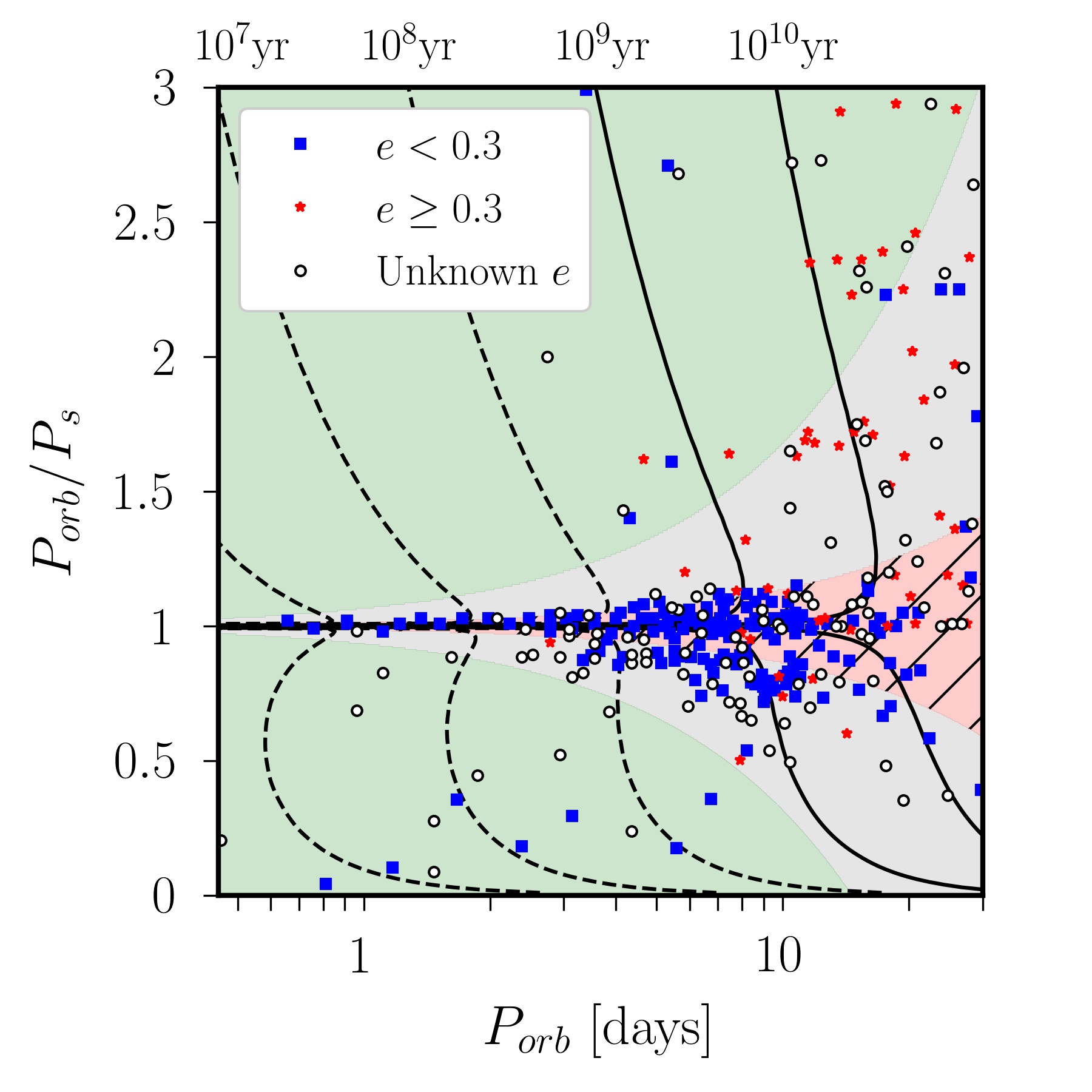} \\
    \includegraphics[width=0.49\textwidth]{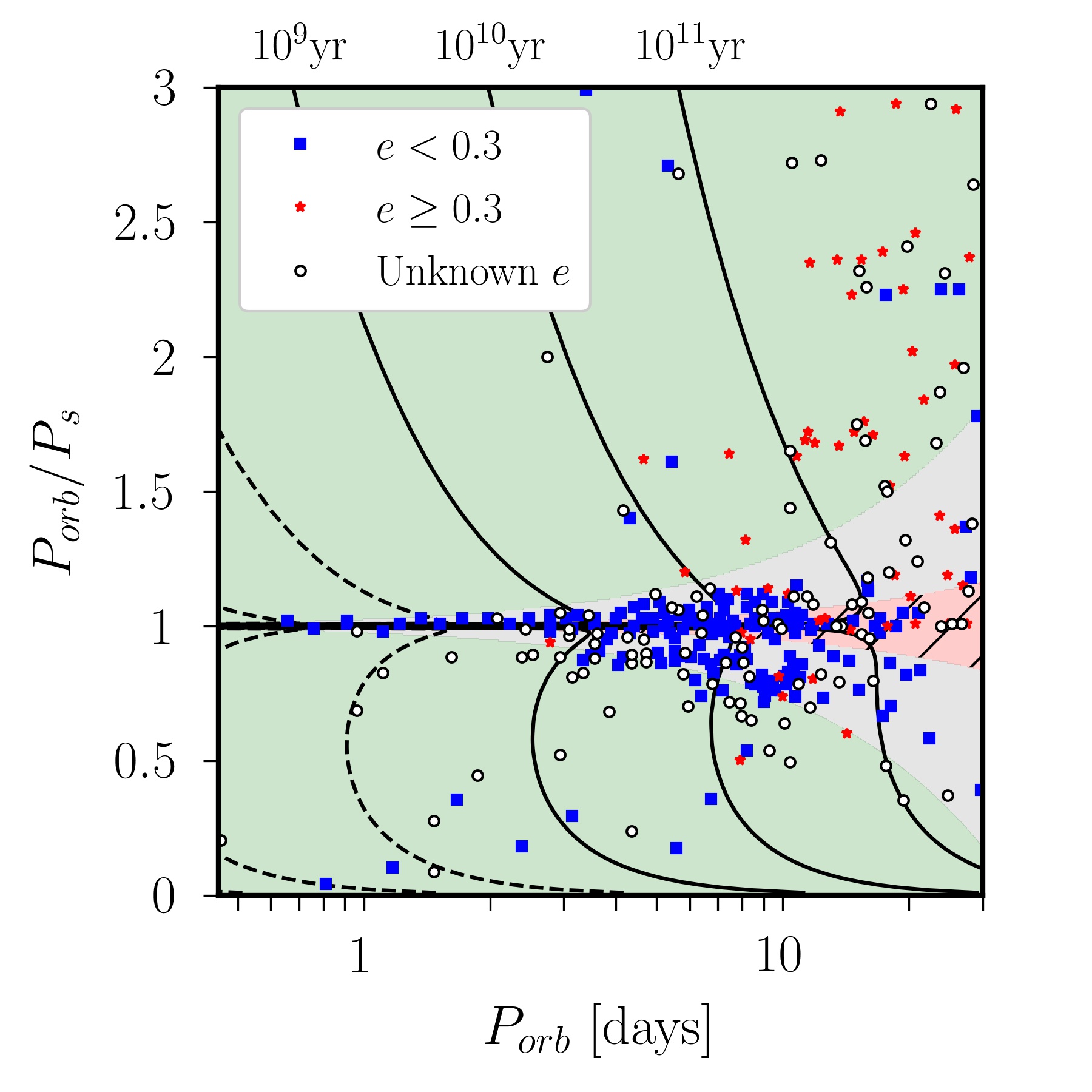} \\
    \end{tabular}
    \caption{Different regimes for the frequency reduction of turbulent viscosity for solar-type binaries, with an assumed transition at $|\omega_t|/\omega_{cv} = 5$.
    Red hashed zone: $\nu_E \simeq \nu_{cv}$. Gray zone: linear reduction. Green zone: quadratic reduction.  
    Binaries, sorted by eccentricity $e$, extracted from Figure 7 in \citet{lurie2017tidal}. \emph{Top}: solar-like star ($1 \, M_\odot$) at $1$ Gyr with $\tau_{cv} \simeq 0.1$ yr and $M_{cv} \approx 0.02 \, M_\odot$.
    \emph{Bottom}: low-mass star ($0.5 \, M_\odot$) at $1$ Gyr with $\tau_{cv} \simeq 0.25$ yr and $M_{cv} \approx 0.3 \, M_\odot$.
   }
    \label{fig:eclipsing}
\end{figure}

We extrapolate our findings to stellar interiors as follows. 
MLT predicts the rms convective velocity to scale as $u_{rms} \propto (Ra/Pr)^{1/2}$ in the fully turbulent regime \citep[e.g.][]{Spiegel1971}, such that $\nu_E/\nu \simeq \nu_{cv}/\nu \propto (Ra/Pr)^{1/2}$ is independent of tidal frequency when $|\omega_t|/\omega_{cv} \ll 1$. 
Such a frequency-independent $\nu_E$ is consistent with constant tidal lag-time models \citep[e.g.][]{hut1981tidal}. 
Then, the effective viscosity is reduced in the presence of fast tides, first with an approximately linear reduction and then a quadratic one. 
The transition between these two regimes occurs when $|\omega_t|/\omega_{cv} \simeq \mathcal{O}(10)$. 
Further work is required to explore the robustness of the transition when $Ra/Pr$ is increased. 
We have also obtained statistically significant negative values of $\nu_E$ for high frequencies in Figure \ref{fig:eddy}, which is consistent with previous local results \citep[][]{ogilvie2012interaction,duguid2019tides}. 
Negative values probably result from (necessarily) adopting simulation parameters that are far removed from their astrophysical values. 
This phenomenon is always observed when $|\nu_E|\leq \nu$ (here in the quadratic regime). 
As also found in \citet{duguid2019tides}, the negative values occur when when $|\omega_t|$ lies in the dissipation range of the turbulence (see Figure \ref{fig:spectrum}). 
MLT predicts that $\omega_{cv} \propto Ra^{1/2}$ in the fully turbulent regime, and that the inertial-like range should extend to higher frequencies. 
Typical values for the Rayleigh and Prandtl numbers in solar-like stars are $Ra=10^{19}-10^{24}$ and $Pr=10^{-6}-10^{-4}$ \citep{hanasoge2014quest}, such that 
we expect $\nu_{cv}\gg \nu$. 
Thus, unrealistically large values of $|\omega_t|$ may be required to get negative values $\nu_E \leq 0$. 

Our results are directly relevant for interpreting observational evidence for synchronization and circularization of solar-type and low-mass close binaries \citep[e.g.][]{meibom2005,meibom2006observational,van2016orbital,lurie2017tidal,triaud2017eblm}. 
We can estimate the convective turnover timescale $\tau_{cv} = 2\pi/\omega_{cv}$ \citep[e.g.][]{terquem1998tidal}, 
using an estimate based on the stellar luminosity \citep[e.g.][]{price2018binary}
\begin{equation}
    \tau_{cv} \approx 0.37 \, \mathrm{yr} \  ({M_{cv}}/{M_\odot})^{1/3} \, ({T_{e}}/{5600})^{-4/3},
    \label{eq:tcvGoodman}
\end{equation}
with $M_{cv}$ the mass of the convective zone, 
$M_\odot$ the solar mass and $T_{e}$ the effective temperature (in Kelvin). 
Assuming a transition between the two regimes when $|\omega_t|/\omega_{cv} \simeq 5$ (see Figure \ref{fig:eddy}), the quadratic scaling should be used when $P_t = 2\pi/|\omega_t| \lesssim 7.3-18.1$ days for a convective timescale $\tau_{cv} \sim 0.1-0.25$ yr, 
where $P_t = |1/P_s - 1/P_{orb}|^{-1} /2$ is the forcing period, $P_s$ the rotation period and $P_{orb}$ the orbital period (in nonsynchronized systems). 
Since low-mass stars typically have longer timescales $\tau_{cv}$, the transition occurs for larger orbital periods for these objects
(see Figure \ref{fig:eclipsing}). 

The range of validity of the various turbulent viscosity prescriptions is shown in Figure \ref{fig:eclipsing}, for binaries given in \citet{lurie2017tidal}. 
Both scalings are shown to be relevant for this sample. 
Therefore, equilibrium tide theory must be carefully applied to interpret the observational data. 
Convective damping of the equilibrium tide could potentially explain the main features of this distribution. 
The timescale for tidal spin-synchronization, for a solar-mass binary in a circular orbit with $P_\mathrm{orb}=10$ days and $P_s = 15$ days, is estimated to be approximately 1 Gyr if we adopt a continuous profile for $\nu_E(|\omega_t|)$ to fit Figure \ref{fig:eddy} (see Appendix \ref{appendix:extrapolation}). 
This seems to be an efficient mechanism for $P_{orb}\lesssim 10$ d. 
We have also superimposed our theoretical predictions for spin-synchronization timescales (due to convective damping of the equilibrium tide), using two different stellar models (see Appendix \ref{appendix:extrapolation}) that span the majority of the sample in \citet{lurie2017tidal}. 
They suggest that the quadratic reduction could explain, for larger values of $|\omega_t|$, why some short-period binaries in Figure \ref{fig:eclipsing} have not yet synchronized. 
Finally, the mechanism seems too efficient to explain why some systems with $P_{orb}<10$ days and $P_{orb}<P_s$ are not synchronized. 
This may be due to the young ages or high masses of these stars, or perhaps because they are affected by differential rotation \citep{lurie2017tidal}. 
The dynamical tide may also be important for some of these systems \citep[e.g.][]{OL2007,ogilvie2014tidal}. 

\section{Concluding remarks}
\label{sec:ccl}
In this Letter, we have revisited the long-standing problem of the interaction between tidal flows and turbulent convection. 
We have conducted the first numerical simulations of turbulent convection within an idealized global model of low-mass fully convective stars (or core-less giant planets), to measure the turbulent viscosity acting on the large-scale equilibrium tidal flow. 
We have reconciled, for the first time and within a single consistent physical model, the two contradictory scaling laws that have been proposed to describe the frequency reduction of the effective viscosity when the tidal frequency exceeds the dominant convective turnover frequency
\citep[i.e. fast tides,][]{zahn1966marees,goldreich1977turbulent}. 
Our results have confirmed the universality of the quadratic reduction in the high-frequency regime, that is $\nu_E\propto |\omega_t|^{-2}$ when $|\omega_t|/\omega_{cv} \gg 1$ \citep[in local models, see][]{ogilvie2012interaction,duguid2019tides}. 
Moreover, we find evidence for a linear reduction ($\nu_E\propto |\omega_t|^{-1}$), in an intermediate regime $1 \lesssim |\omega_t|/\omega_{cv} \lesssim \mathcal{O}(10)$.
This likely results from the non-Kolmogorov nature of the turbulence in that frequency range \citep[e.g.][]{penev2007dissipation,penev2009dissipation,penev2009direct}. 
This has important consequences for interpreting astrophysical observations.
Our findings should guide future data-driven studies to discriminate between these two scaling laws, for instance, when interpreting observations of the 
synchronization and circularization of main-sequence binaries \citep{lurie2017tidal,triaud2017eblm} or the circularization of evolved stars 
\citep{price2018binary}.

Much further work is required before we can accurately model the tidal evolution of astrophysical systems due to this mechanism. 
The robustness and coexistence of these two scaling regimes, and the transition between them, should be explored further. 
Moreover, we have neglected dynamical tides \citep[e.g.][]{OL2007} and considered only circular orbits. 
Different tidal components could, however, be damped at different rates \citep[e.g.][]{lai2012tidal}. 
Simulations in spherical shells would be also worth exploring \citep[e.g.][]{gastine2016scaling} to model the convective envelopes of solar-like stars.
Given the importance of this problem, understanding the interaction between turbulent convection and tidal flows appears urgent.
This is necessary to correctly interpret observations of close binaries 
\citep[e.g.][]{lurie2017tidal,triaud2017eblm,price2018binary}. 

The validity of MLT should be also assessed using turbulent simulations of convection. 
MLT predictions could underestimate the turbulent viscosity $\nu_{cv}$ in the low-frequency regime \citep{goldman2008effective}. Indeed, departures from MLT have been found in recent simulations of compressible convection \citep[e.g.][]{anders2019entropy}.
Moreover, convection-driven turbulence is strongly affected by rapid rotation \citep[e.g.][]{gastine2016scaling,kaplan2017subcritical}, such as in giant planets or young stars.
The prescription for the turbulent viscosity from MLT \citep[e.g.][]{barker2014theory} then ought to be modified \citep[see][in the low-frequency regime]{mathis2016impact}. 
More realistic convection models should be considered as a long-term endeavor.
Finally, by neglecting rotation, we have also filtered out nonlinear tidal flows such as the elliptical (tidal) instability \citep[e.g.][]{barker2016non,vidal2017inviscid}. 
They could enhance tidal dissipation for the shortest orbital periods
\citep{barker2016nonb,vidal2018magnetic,vidal2019binaries},
and might even modify properties of turbulent convection \citep{cebron2010tidal}. 
Understanding their interplay with convection deserves future work.

\section*{Acknowledgments}
This work was funded by STFC grant ST/R00059X/1.
JV is grateful to S. Reddy for his support with Nek5000. 
Numerical simulations were undertaken on ARC1/ARC2 clusters (HPC Facility, University of Leeds).
Simulations were also performed on the DiRAC Data Intensive service at Leicester (STFC DiRAC HPC Facility). 
The equipment was funded by BEIS capital funding via 
STFC capital grants ST/K000373/1 and ST/R002363/1 and STFC DiRAC Operations grant ST/R001014/1.
This work also used the DiRAC@Durham facility.
The equipment was funded 
by BEIS capital funding via 
STFC capital grants ST/P002293/1, ST/R002371/1 and ST/S002502/1, Durham University and STFC operations grant ST/R000832/1. 
DiRAC is part of the National e-Infrastructure.
We acknowledge the anonymous referee for several suggestions that have allowed us to improve the paper.




\software{Nek5000 (\url{https://nek5000.mcs.anl.gov/}).}

\appendix
\section{Convection Model}
\label{appendix:convection}
We study Boussinesq thermal convection \citep[e.g.][]{Spiegel1971}, driven by homogeneous internal heating $\mathcal{Q}_T$ in a full sphere. 
We use the notation introduced in Section \ref{sec:model}. 
The gravitational field is $\boldsymbol{g} = -g_0 \, \boldsymbol{r}$, where $\boldsymbol{r}$ is the position vector and $g_0$ is a constant. 
This is the leading-order component for a low-mass body that is not very centrally condensed \citep[e.g.][]{lai1993ellipsoidal}. 
We employ dimensionless quantities for the simulations, adopting $R$ as the length scale, the viscous timescale ${R^2}/{\nu}$ as the timescale, and $(\nu \mathcal{Q}_T R^2)/(6 {\kappa}^2)$ as the unit of temperature \citep[as in][]{monville2019rotating}. 
The dimensionless equations for $\boldsymbol{u}$ and the temperature perturbation $\Theta$, in the inertial frame, are
\begin{subequations}
\allowdisplaybreaks
  \label{eqModeling:GeneralEquations}
	\begin{align}
    \frac{\partial \boldsymbol{u}}{\partial t} + (\boldsymbol{u} \boldsymbol{\cdot} \boldsymbol{\nabla} ) \,  \boldsymbol{u} &= 
    - \nabla p +  \boldsymbol{\nabla}^2 \boldsymbol{u} + Ra \, \Theta \, \boldsymbol{r} - \boldsymbol{f}, \\
    \frac{\partial \Theta}{\partial t} + (\boldsymbol{u} \boldsymbol{\cdot} \nabla) \, \Theta &=  \frac{1}{Pr} \left [ 2 \, \boldsymbol{u} \boldsymbol{\cdot} \boldsymbol{r} + \nabla^2 \Theta \right ] - \mathcal{Q},
    \end{align}
\end{subequations}
where $p$ is a dimensionless (reduced) pressure, $\boldsymbol{f} = ( \boldsymbol{u} \boldsymbol{\cdot} \boldsymbol{\nabla} ) \, \boldsymbol{U}_0 + ( \boldsymbol{U}_0 \boldsymbol{\cdot} \boldsymbol{\nabla} ) \, \boldsymbol{u}$ a forcing term with $\boldsymbol{U}_0$ given by (\ref{eq:U0}) and $\mathcal{Q} = (\boldsymbol{U}_0 \boldsymbol{\cdot} \nabla) \, \Theta$. 
We have defined the Rayleigh number 
$Ra = {\alpha_T g_0 \mathcal{Q}_T R^6}/(6\nu \kappa^2)$, where $\alpha_T$ is the thermal expansion coefficient, and the Prandtl number $Pr={\nu}/{\kappa}$. 
The nonlinear term $(\boldsymbol{U}_0 \boldsymbol{\cdot}  \boldsymbol{\nabla}) \, \boldsymbol{U}_0$ reduces here to a pressure gradient, and thus plays no dynamical role within the Boussinesq approximation. 
We have also neglected in $\mathcal{Q}$ the term $(\boldsymbol{U}_0 \boldsymbol{\cdot} \nabla) \, T_0$ that should vanish in the limit $\beta \ll 1$ \citep[e.g.][in the ellipsoidal geometry]{lai1993ellipsoidal}, where $T_0$ is the background temperature.  
Equations (\ref{eqModeling:GeneralEquations}) are complemented with the incompressibility condition $  \boldsymbol{\nabla} \boldsymbol{\cdot} \boldsymbol{u} = 0$, and boundary conditions at the (dimensionless) spherical boundary $r=1$. 
For the temperature, we employ the isothermal condition $\Theta=0$. 
To avoid spurious numerical issues associated with angular momentum conservation in global simulations of tidal flows \citep[as explained in][]{guermond2013remarks,favier2014non}, we enforce the mechanical boundary condition $\boldsymbol{u}=\boldsymbol{0}$. 
This is unlikely to affect the (small-scale) turbulent flows driven in the bulk without rotation (compared to stress-free boundary conditions). 

We have solved nonlinear Equations (\ref{eqModeling:GeneralEquations}) in their weak variational form by using the spectral-element code Nek5000 \citep[e.g.][]{fischer2007simulation}. 
The computational domain is decomposed into $E=3584$ non-overlapping hexahedral elements. 
Within each element, the velocity (and pressure) is represented as Lagrange polynomials of order $N$ (respectively, $N-2$) on the Gauss-Lobatto-Legendre (Gauss-Legendre) points. 
Temporal discretization is accomplished by a third-order method, based on an adaptive and semi-implicit scheme in which the nonlinear and Coriolis terms are treated explicitly, and the remaining linear terms are treated implicitly. 
Solutions are de-aliased following the $3/2$ rule, such that $3N/2$ grid points are used in each dimension for the nonlinear terms, whereas only $N$ points are used for the linear terms.
We have checked the numerical accuracy in targeted simulations to ensure convergence by varying the polynomial order from $N=7$ to $N=9$. We adopt a time step $10^{-6} \leq \mathrm{d} t \leq 5 \times 10^{-6}$ (in dimensionless units, depending on the forcing frequency). 

For most of the simulations, we initiated the convection with random noise to the temperature field and let it saturate without tides (i.e. $\beta=0$), before switching on the equilibrium tidal flow. 
We have checked that initiating the convection together with the equilibrium tidal flow does not lead to noticeably different results. 
We have integrated each simulation for several viscous timescales ($5 \leq \Delta T \leq 10$ in dimensionless units), corresponding with more than a hundred tidal periods, to obtain converged statistics for the effective viscosity. 
The time average in expression (\ref{eq:eddy}) is obtained by fitting a linear slope to the cumulative time integral 
\citep[e.g. see Figure 13 in][]{duguid2019tides}, to reduce the turbulent noise. 

\section{Complementary Results}
\label{appendix:results}
The parameters and results of the simulations behind Figure \ref{fig:eddy} are given in Table \ref{tab:beta15e-2}. 
We define the rms velocity $u_{rms}$ as the time average of $(2 \, \langle\mathcal{E}(\boldsymbol{u})\rangle_V /3)^{1/2}$ with $\mathcal{E}(\boldsymbol{u}) = (u_x^2 + u_y^2 + u_z^2 )/2$ the kinetic energy, noting that there is no preferred Cartesian direction for the flow without rotation. 
The turbulent length scale is estimated (by eye) as $l_E \simeq 1/3$, which agrees with Figure \ref{fig:convection}. 
The latter figure indeed shows that multiple eddies span the radius of the body. 
Then, we define the turnover frequency as $\omega_{cv}=u_{rms}/l_E$. 
Small differences in the rms properties of the convection are found when the amplitude of the tidal flow was larger than the convective flow (i.e. when $\beta |\omega_t| \geq u_{rms}$). 
Typically, these differences are smaller than $5\%$ for the kinetic energy and the rms velocity when $\beta \leq 5\times 10^{-2}$. However, in the strong tides regime the convection can be modified more significantly, which we have observed when $\beta\geq 10^{-1}$, or for very high frequencies (i.e. $|\omega_t|/ \omega_{cv} \gg 100$ at $Ra=10^6$ with $\beta=5\times10^{-2}$). Similar findings have been reported in local simulations \citep[e.g.][]{duguid2019tides}.

\begin{figure}
    \centering
    \begin{tabular}{c}
    \includegraphics[width=0.49\textwidth]{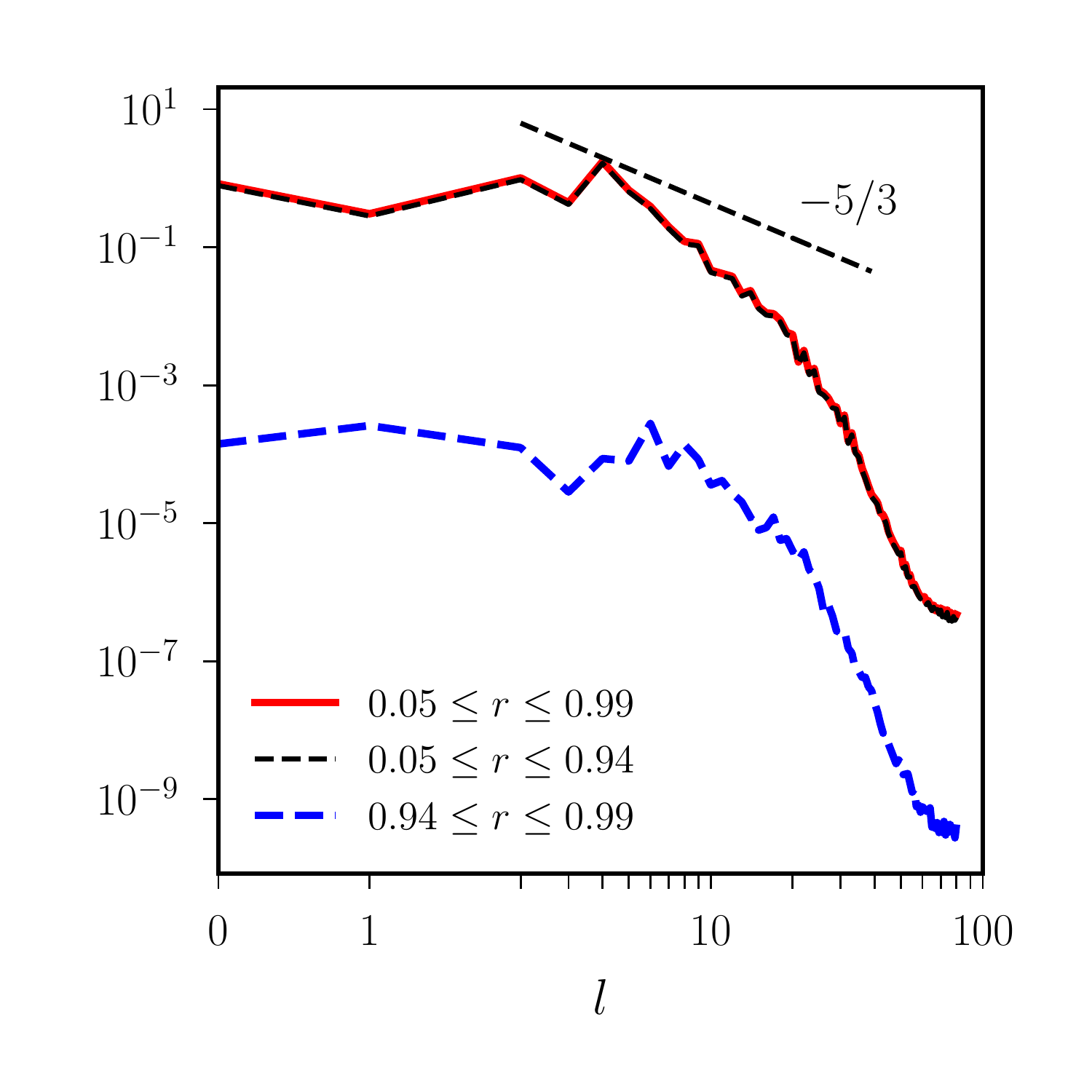} \\
    \includegraphics[width=0.49\textwidth]{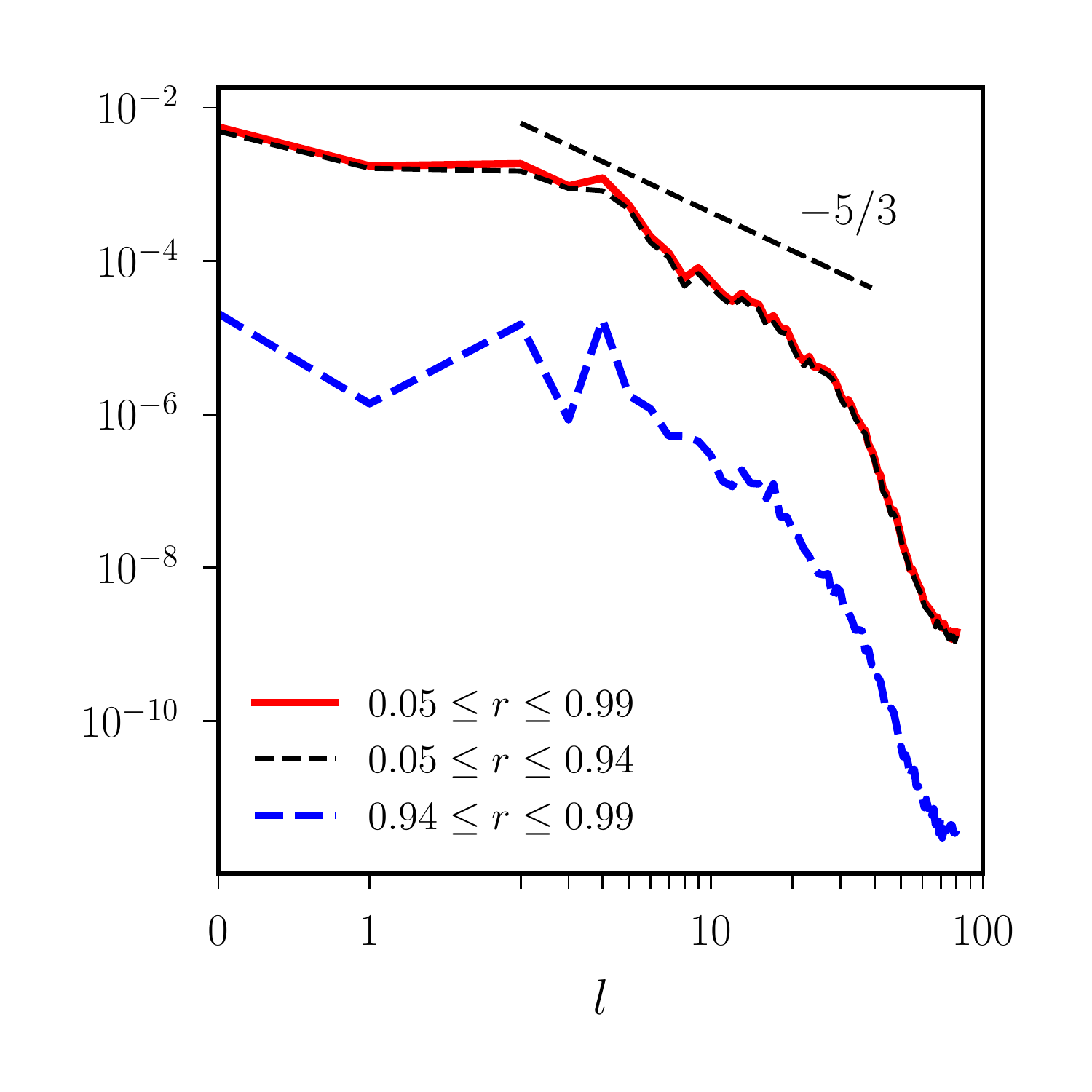} \\
    \end{tabular}
    \caption{Power spectrum of the time-averaged and radially integrated quantity 
    $-{1}/{(\Delta T (\omega_t \beta)^2)} \, \int \boldsymbol{u} \boldsymbol{\cdot} [\boldsymbol{u} \boldsymbol{\cdot} \boldsymbol{\nabla}\boldsymbol{U}_0] \, \mathrm{d} t$, as a function of the spherical harmonic degree $l\geq 0$ (using orthonormalized harmonics). Simulations at $Ra=10^6, Pr=1$ and $\beta=5\times10^{-2}$. 
    Eddy viscosity (\ref{eq:eddy}) is given by the square root of the $l=0$ component (in the physical space), when $\Delta T$ is large enough to reduce the turbulent noise. \emph{Top}: $|\omega_t|/\omega_{cv} = 2.3$ with an integration time $\Delta T \simeq 0.59$. \emph{Bottom}: $|\omega_t|/\omega_{cv} = 13$ with an integration time $\Delta T \simeq 0.39$. 
    Spectra have been computed by interpolating the data to a spherical grid. 
    }
    \label{fig:eddyBL}
\end{figure}

We illustrate in Figure \ref{fig:eddyBL} spatial spectra of the term $\boldsymbol{u} \boldsymbol{\cdot} [\boldsymbol{u} \boldsymbol{\cdot} \boldsymbol{\nabla}\boldsymbol{U}_0]$ that appears in equation (\ref{eq:eddy}) for the effective viscosity. We have computed it in the entire fluid domain (i.e. $0.05\leq r \leq0.99$) and omitting the boundary regions (i.e. $0.94\leq r \leq0.99$). We find that the eddy viscosity (i.e. the $l=0$ component in the physical space) is never dominated by interactions near the boundary, but is instead due to flows in the bulk.

\begin{table*}[t]
	\centering
    \caption{Table of Simulation Results for $Ra=10^6$ and $Pr=1$.}
    \label{tab:beta15e-2}
    \begin{tabular}{ccccccc}
        \hline
        \hline
        $|\omega_t|$ & $\langle \mathcal{E}(\boldsymbol{u}) \rangle_V$ & $u_{rms}$ & $|\omega_t|/\omega_{cv}$ & $\nu_E$ & UpEb & LwEb \\
        \hline
$2.0 \times 10^{+2}$ & $3.6 \times 10^{+3}$ & $4.9 \times 10^{+1}$ & $1.4 \times 10^{+0}$ & $+1.2 \times 10^{+1}$ & $1.1 \times 10^{+1}$ & $6.9 \times 10^{+0}$ \\
$2.2 \times 10^{+2}$ & $3.6 \times 10^{+3}$ & $4.9 \times 10^{+1}$ & $1.5 \times 10^{+0}$ & $+1.5 \times 10^{+1}$ & $1.1 \times 10^{+1}$ & $8.0 \times 10^{+0}$ \\
$2.5 \times 10^{+2}$ & $3.6 \times 10^{+3}$ & $4.9 \times 10^{+1}$ & $1.7 \times 10^{+0}$ & $+1.2 \times 10^{+1}$ & $8.3 \times 10^{+0}$ & $5.4 \times 10^{+0}$ \\
$2.9 \times 10^{+2}$ & $3.5 \times 10^{+3}$ & $4.8 \times 10^{+1}$ & $2.0 \times 10^{+0}$ & $+1.4 \times 10^{+1}$ & $3.8 \times 10^{+0}$ & $3.8 \times 10^{+0}$ \\
$3.3 \times 10^{+2}$ & $3.6 \times 10^{+3}$ & $4.9 \times 10^{+1}$ & $2.3 \times 10^{+0}$ & $+1.1 \times 10^{+1}$ & $6.4 \times 10^{+0}$ & $4.8 \times 10^{+0}$ \\
$4.0 \times 10^{+2}$ & $3.6 \times 10^{+3}$ & $4.9 \times 10^{+1}$ & $2.7 \times 10^{+0}$ & $+7.8 \times 10^{+0}$ & $2.1 \times 10^{+0}$ & $2.1 \times 10^{+0}$ \\
$5.0 \times 10^{+2}$ & $3.6 \times 10^{+3}$ & $4.9 \times 10^{+1}$ & $3.4 \times 10^{+0}$ & $+3.5 \times 10^{+0}$ & $1.1 \times 10^{+0}$ & $1.1 \times 10^{+0}$ \\
$5.7 \times 10^{+2}$ & $3.6 \times 10^{+3}$ & $4.9 \times 10^{+1}$ & $3.9 \times 10^{+0}$ & $+2.4 \times 10^{+0}$ & $9.6 \times 10^{-1}$ & $9.6 \times 10^{-1}$ \\
$6.7 \times 10^{+2}$ & $3.6 \times 10^{+3}$ & $4.9 \times 10^{+1}$ & $4.6 \times 10^{+0}$ & $+1.8 \times 10^{+0}$ & $1.2 \times 10^{+0}$ & $9.1 \times 10^{-1}$ \\
$8.0 \times 10^{+2}$ & $3.6 \times 10^{+3}$ & $4.9 \times 10^{+1}$ & $5.5 \times 10^{+0}$ & $+9.7 \times 10^{-1}$ & $8.1 \times 10^{-1}$ & $6.1 \times 10^{-1}$ \\
$1.0 \times 10^{+3}$ & $3.6 \times 10^{+3}$ & $4.9 \times 10^{+1}$ & $6.8 \times 10^{+0}$ & $+4.0 \times 10^{-1}$ & $3.2 \times 10^{-1}$ & $2.4 \times 10^{-1}$ \\
$1.3 \times 10^{+3}$ & $3.7 \times 10^{+3}$ & $4.9 \times 10^{+1}$ & $9.0 \times 10^{+0}$ & $-1.4 \times 10^{-1}$ & $7.8 \times 10^{-2}$ & $7.8 \times 10^{-2}$ \\
$2.0 \times 10^{+3}$ & $3.6 \times 10^{+3}$ & $4.9 \times 10^{+1}$ & $1.4 \times 10^{+1}$ & $-1.2 \times 10^{-1}$ & $7.3 \times 10^{-2}$ & $5.5 \times 10^{-2}$ \\
$3.3 \times 10^{+3}$ & $3.6 \times 10^{+3}$ & $4.9 \times 10^{+1}$ & $2.3 \times 10^{+1}$ & $-3.8 \times 10^{-2}$ & $2.5 \times 10^{-2}$ & $1.9 \times 10^{-2}$ \\
$6.7 \times 10^{+3}$ & $3.6 \times 10^{+3}$ & $4.9 \times 10^{+1}$ & $4.5 \times 10^{+1}$ & $-1.8 \times 10^{-2}$ & $1.4 \times 10^{-2}$ & $1.1 \times 10^{-2}$ \\
$1.0 \times 10^{+4}$ & $3.6 \times 10^{+3}$ & $4.9 \times 10^{+1}$ & $6.8 \times 10^{+1}$ & $-1.1 \times 10^{-2}$ & $1.7 \times 10^{-3}$ & $4.4 \times 10^{-3}$ \\
$2.0 \times 10^{+4}$ & $3.6 \times 10^{+3}$ & $4.9 \times 10^{+1}$ & $1.4 \times 10^{+2}$ & $-3.7 \times 10^{-3}$ & $1.2 \times 10^{-3}$ & $1.2 \times 10^{-3}$ \\
\hline
    \end{tabular}
    \vspace{1em}

    \centering
    \begin{tabular}{ccccccc}
        \hline
        $|\omega_t|$ & $\langle \mathcal{E}(\boldsymbol{u})\rangle_V$ & $u_{rms}$ & $|\omega_t|/\omega_{cv}$ & $\nu_E$ & UpEb & LwEb \\
        \hline
$2.0 \times 10^{+2}$ & $3.3 \times 10^{+3}$ & $4.7 \times 10^{+1}$ & $1.4 \times 10^{+0}$ & $+1.0 \times 10^{+1}$ & $4.0 \times 10^{+0}$ & $2.7 \times 10^{+0}$ \\
$2.2 \times 10^{+2}$ & $3.3 \times 10^{+3}$ & $4.7 \times 10^{+1}$ & $1.6 \times 10^{+0}$ & $+1.2 \times 10^{+1}$ & $1.8 \times 10^{+0}$ & $2.7 \times 10^{+0}$ \\
$2.5 \times 10^{+2}$ & $3.3 \times 10^{+3}$ & $4.7 \times 10^{+1}$ & $1.8 \times 10^{+0}$ & $+1.2 \times 10^{+1}$ & $1.3 \times 10^{+0}$ & $1.9 \times 10^{+0}$ \\
$2.9 \times 10^{+2}$ & $3.2 \times 10^{+3}$ & $4.6 \times 10^{+1}$ & $2.0 \times 10^{+0}$ & $+1.0 \times 10^{+1}$ & $1.7 \times 10^{+0}$ & $2.6 \times 10^{+0}$ \\
$3.3 \times 10^{+2}$ & $3.1 \times 10^{+3}$ & $4.6 \times 10^{+1}$ & $2.3 \times 10^{+0}$ & $+8.7 \times 10^{+0}$ & $1.2 \times 10^{+0}$ & $8.0 \times 10^{-1}$ \\
$4.0 \times 10^{+2}$ & $3.1 \times 10^{+3}$ & $4.6 \times 10^{+1}$ & $2.7 \times 10^{+0}$ & $+7.3 \times 10^{+0}$ & $5.4 \times 10^{-1}$ & $1.1 \times 10^{+0}$ \\
$5.0 \times 10^{+2}$ & $3.2 \times 10^{+3}$ & $4.6 \times 10^{+1}$ & $3.8 \times 10^{+0}$ & $+6.8 \times 10^{+0}$ & $6.9 \times 10^{-1}$ & $1.0 \times 10^{+0}$ \\
$6.7 \times 10^{+2}$ & $3.4 \times 10^{+3}$ & $4.7 \times 10^{+1}$ & $4.3 \times 10^{+0}$ & $+4.8 \times 10^{+0}$ & $1.4 \times 10^{+0}$ & $1.4 \times 10^{+0}$ \\
$8.0 \times 10^{+2}$ & $3.5 \times 10^{+3}$ & $4.8 \times 10^{+1}$ & $5.9 \times 10^{+0}$ & $+3.8 \times 10^{+0}$ & $1.8 \times 10^{+0}$ & $1.5 \times 10^{+0}$ \\
$9.1 \times 10^{+2}$ & $3.5 \times 10^{+3}$ & $4.8 \times 10^{+1}$ & $6.8 \times 10^{+0}$ & $+3.2 \times 10^{+0}$ & $3.1 \times 10^{-1}$ & $6.1 \times 10^{-1}$ \\
$1.0 \times 10^{+3}$ & $3.5 \times 10^{+3}$ & $4.9 \times 10^{+1}$ & $7.1 \times 10^{+0}$ & $+5.4 \times 10^{-1}$ & $3.2 \times 10^{-1}$ & $3.2 \times 10^{-1}$ \\
$1.1 \times 10^{+3}$ & $3.5 \times 10^{+3}$ & $4.8 \times 10^{+1}$ & $7.7 \times 10^{+0}$ & $+2.3 \times 10^{-1}$ & $1.1 \times 10^{-1}$ & $1.4 \times 10^{-1}$ \\
$1.3 \times 10^{+3}$ & $3.5 \times 10^{+3}$ & $4.9 \times 10^{+1}$ & $9.1 \times 10^{+0}$ & $-1.3 \times 10^{-1}$ & $3.5 \times 10^{-2}$ & $5.2 \times 10^{-2}$ \\
$2.0 \times 10^{+3}$ & $3.5 \times 10^{+3}$ & $4.8 \times 10^{+1}$ & $1.3 \times 10^{+1}$ & $-9.0 \times 10^{-2}$ & $3.4 \times 10^{-2}$ & $3.4 \times 10^{-2}$ \\
$3.3 \times 10^{+3}$ & $3.5 \times 10^{+3}$ & $4.8 \times 10^{+1}$ & $2.3 \times 10^{+1}$ & $-5.1 \times 10^{-2}$ & $2.6 \times 10^{-2}$ & $1.7 \times 10^{-2}$ \\
$6.7 \times 10^{+3}$ & $3.4 \times 10^{+3}$ & $4.8 \times 10^{+1}$ & $4.4 \times 10^{+1}$ & $-1.2 \times 10^{-2}$ & $3.9 \times 10^{-3}$ & $5.8 \times 10^{-3}$ \\
$2.0 \times 10^{+4}$ & $3.4 \times 10^{+3}$ & $4.8 \times 10^{+1}$ & $1.4 \times 10^{+2}$ & $-2.1 \times 10^{-3}$ & $5.1 \times 10^{-4}$ & $7.6 \times 10^{-4}$ \\
        \hline
    \end{tabular}
    \tablecomments{$\langle \mathcal{E}(\boldsymbol{u})\rangle_V$ is the volume-averaged kinetic energy and $u_{rms}$ the rms velocity. UpEb: Upper Error bar. LwEb: Lower Error bar. \emph{Top}: $\beta=10^{-2}$. \emph{Bottom}: $\beta=5\times10^{-2}$.}
\end{table*}

\begin{figure*}[t]
    \centering
    \begin{tabular}{cc}
    \includegraphics[width=0.49\textwidth]{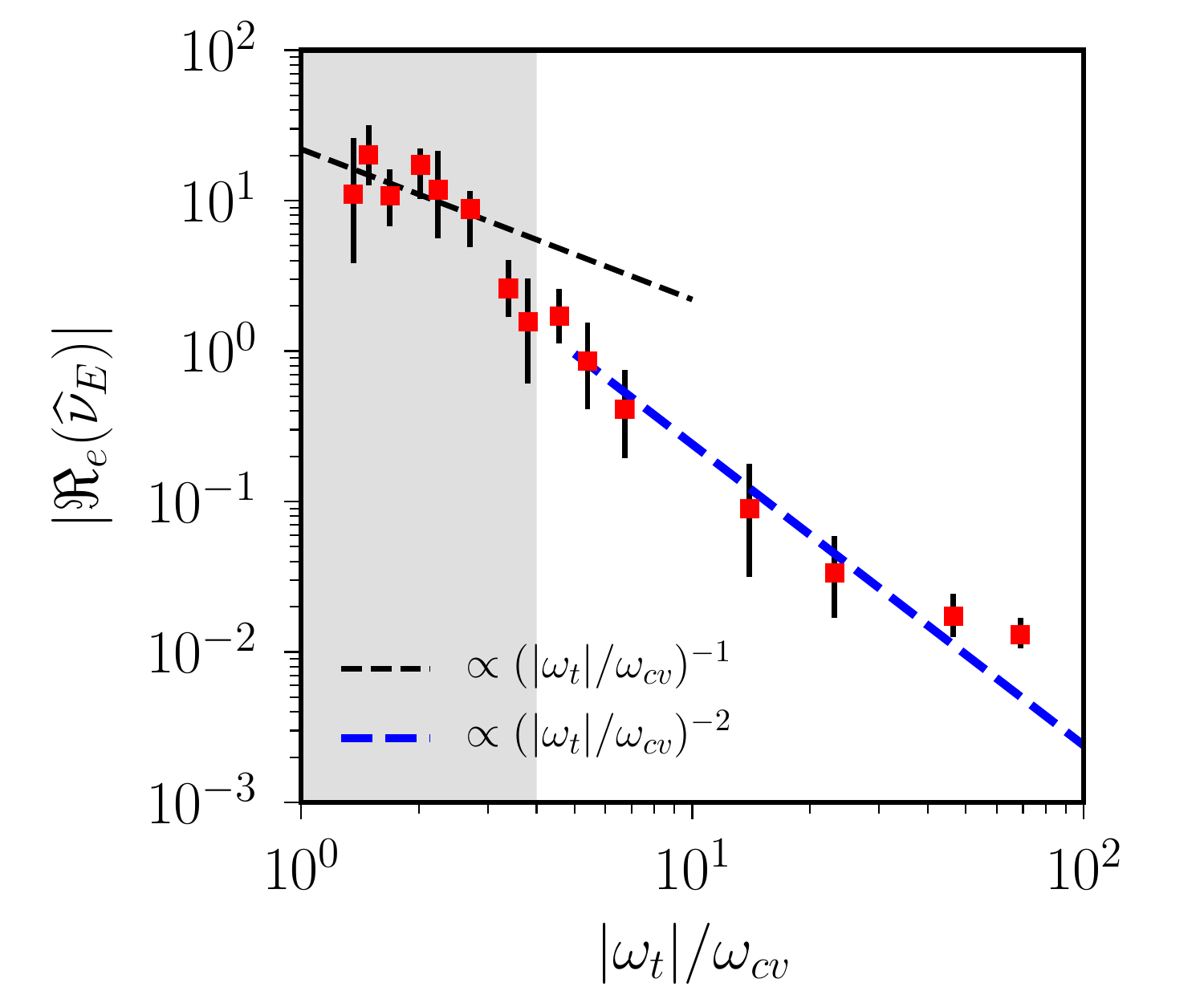} &
    \includegraphics[width=0.49\textwidth]{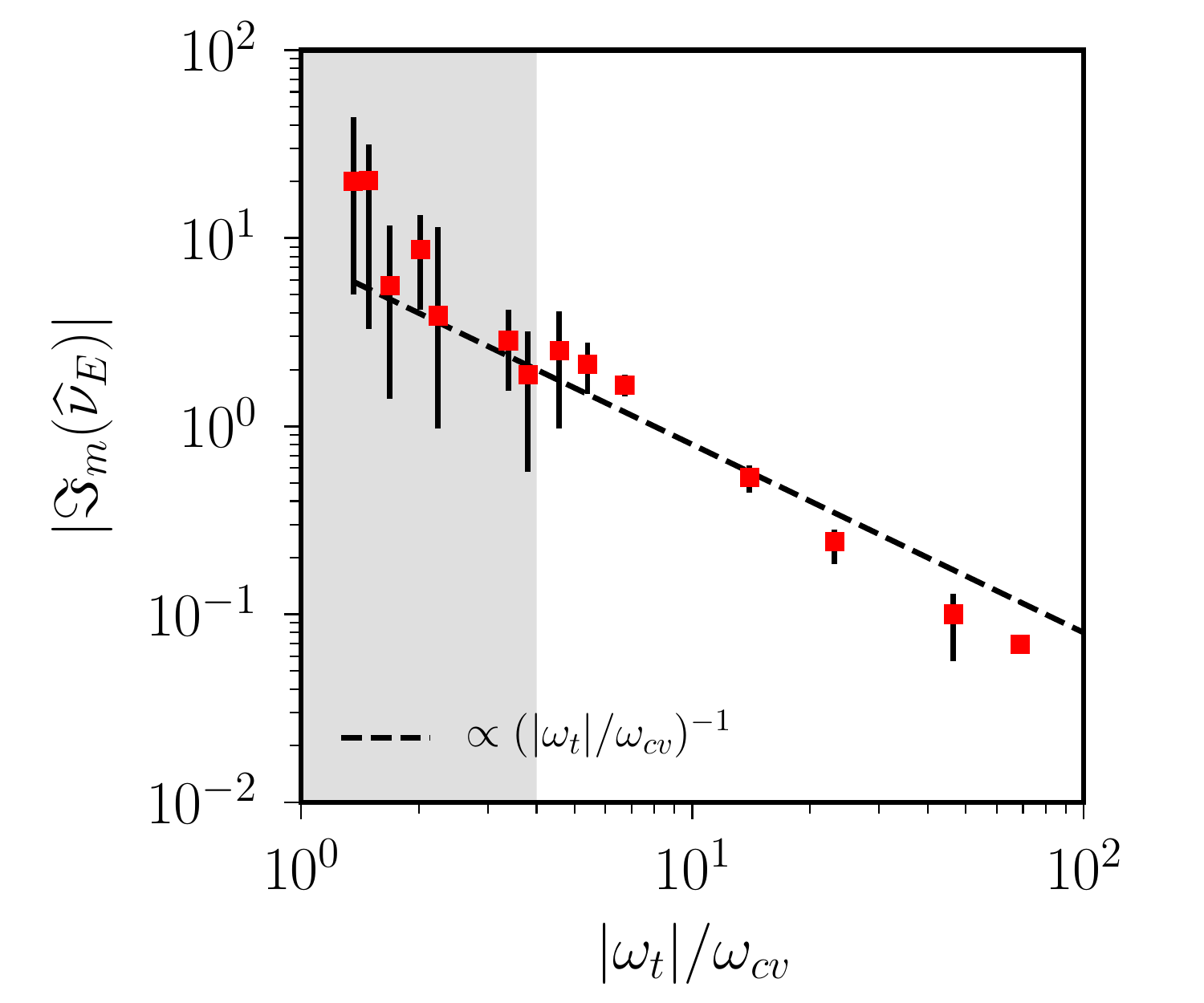} \\
    \includegraphics[width=0.49\textwidth]{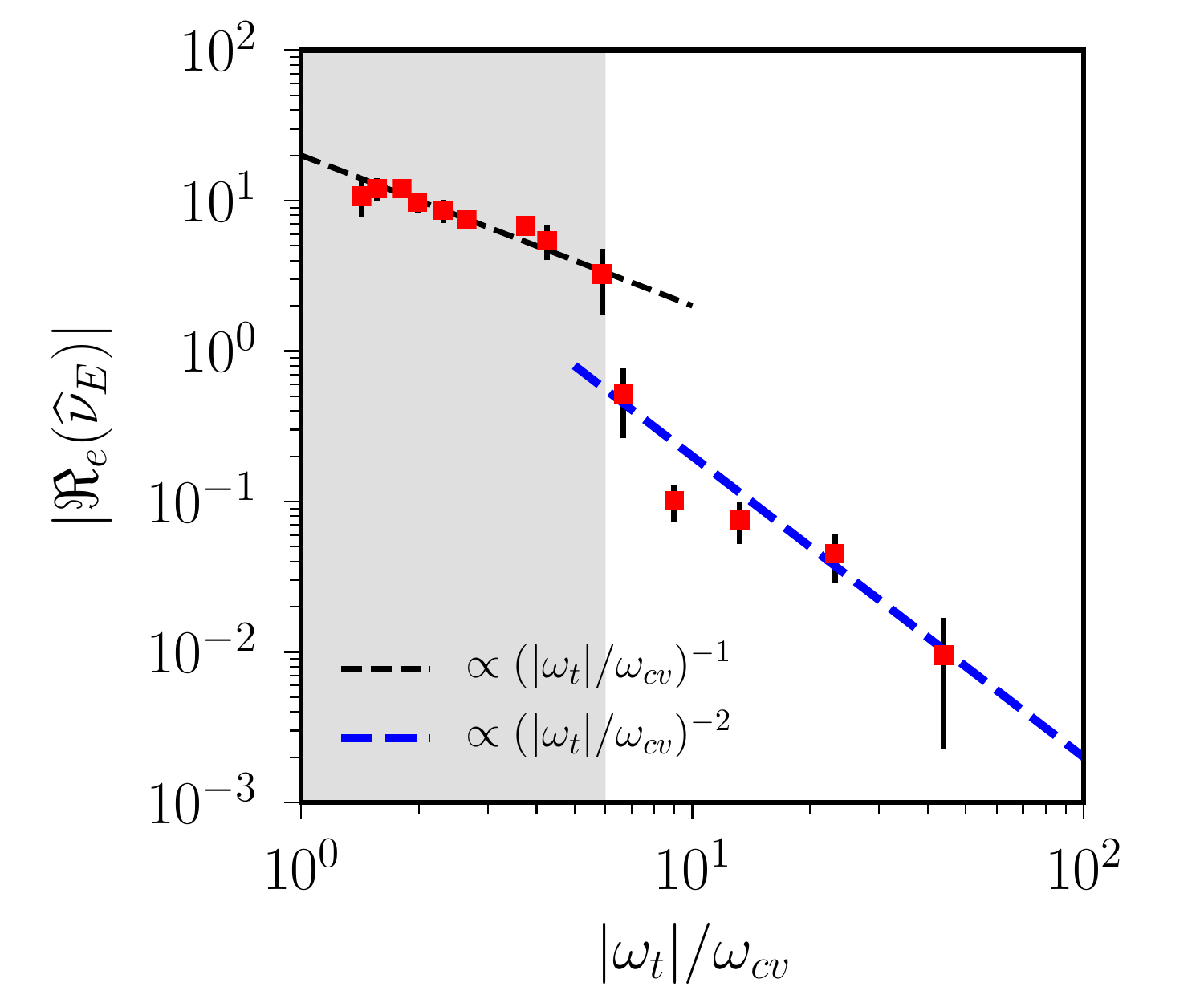} &
    \includegraphics[width=0.49\textwidth]{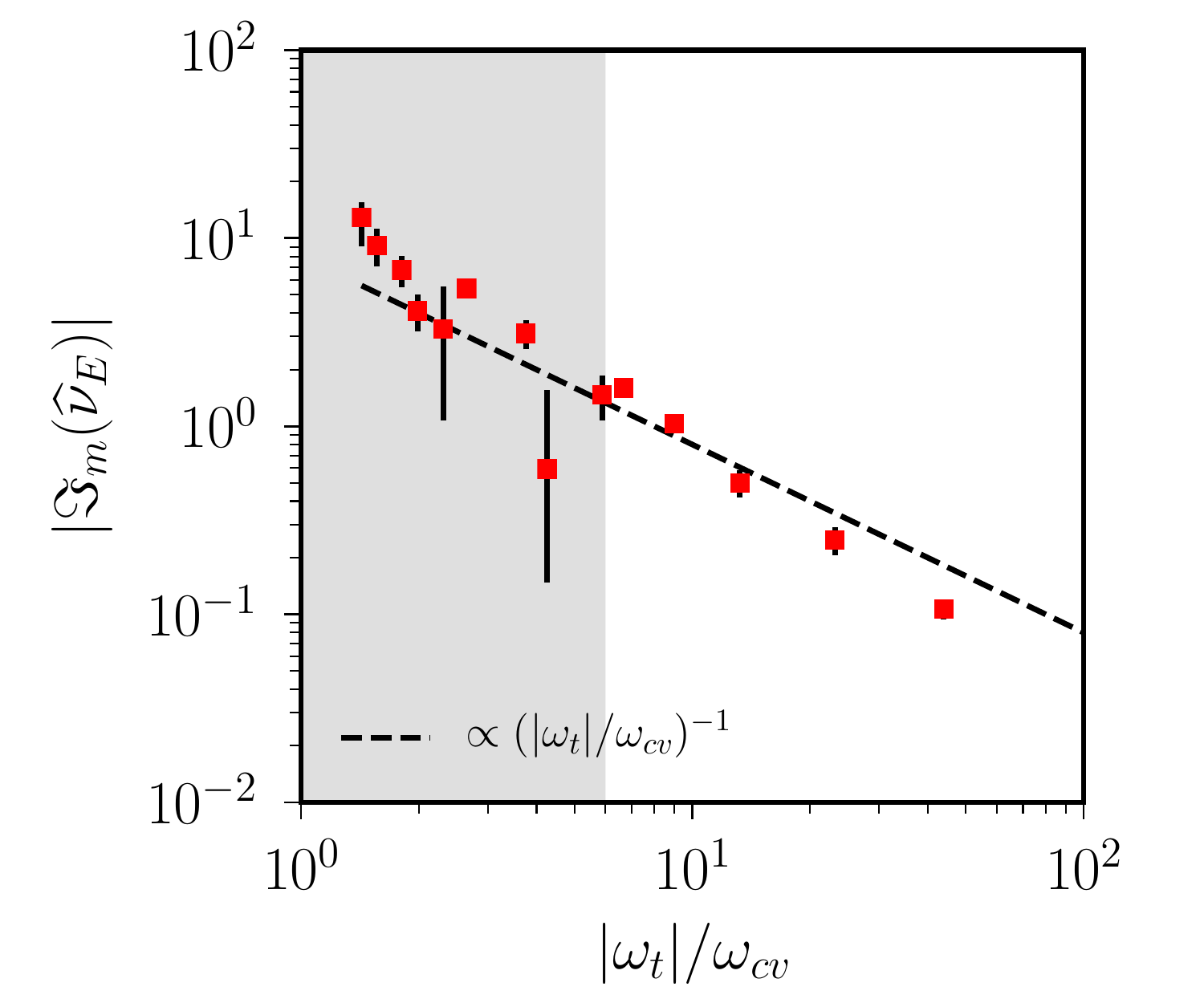} \\
    \end{tabular}
    \caption{Real and imaginary parts of the effective viscosity $|\Re_e(\widehat{\nu}_E)|$ and $|\Im_m(\widehat{\nu}_E)|$, measured from the Reynolds stress tensor, as a function of $|\omega_t|/\omega_{cv}$ in supercritical simulations of convection ($Ra = 10^6 ,Pr=1$). 
    Error bars are computed by evaluating the noise level in the vicinity of the spike at the forcing frequency. 
    The turbulent viscosity (relevant for tidal dissipation) is measured by $|\Re_e(\widehat{\nu}_E)|$, whereas $|\Im_m(\widehat{\nu}_E)|$ measures the elastic component of the response. \emph{Top}: $\beta= 10^{-2}$. \emph{Bottom}: $\beta= 5 \times 10^{-2}$.
    }
    \label{fig:reynoldseddy}
\end{figure*}

For an alternative approach to directly using equation (\ref{eq:eddy}), we can estimate $\nu_E$ by considering the Fourier transform of the volume-averaged component $\langle u_x u_y \rangle_V$, as long as we account for the oscillatory nature of the tidal flow \citep{ogilvie2012interaction}. 
Thus, we can define the effective viscosity $\widehat{\nu}_E$ in the Fourier domain as
\begin{equation}
    \mathcal{F} \{\langle u_x u_y \rangle_V \} = \widehat{\nu}_E \, \omega_t \beta \, \mathcal{F} \{\cos(\omega_t t) \}, \label{eq:closurexy} \\
\end{equation}
where $\mathcal{F}$ denotes the Fourier transform and $\widehat{\nu}_E$ is a complex-valued quantity. 
The Reynolds stress and the rate of strain are generally out of phase.
The real part $\Re_e(\widehat{\nu}_E)\simeq \nu_E$ is the turbulent viscosity (that is in phase with the equilibrium tidal flow), whereas the imaginary part $\Im_m(\widehat{\nu}_E)$ (that is out of phase with the tidal flow) is related to an effective elasticity.
In the regime of high-frequency tidal forcing ($|\omega_t| \gg \omega_{cv}$),
\citet{ogilvie2012interaction} and \citet{duguid2019tides} demonstrated the viscoelastic nature of the tidal response.  
They predict the turbulent viscosity $\Re_e(\widehat{\nu}_E)$ should scale as $|\omega_t|^{-2}$, whereas $\Im_m(\widehat{\nu}_E)$ should obey a linear reduction $|\omega_t|^{-1}$.  

We compute $\widehat{\nu}_E$ from expression (\ref{eq:closurexy}) in Figure \ref{fig:reynoldseddy}. 
The results confirm the universal nature of the viscoelastic response, with a dominant elastic component at high frequencies.  
Indeed, we broadly obtain a linear reduction $|\Im_m(\widehat{\nu}_E)| \propto |\omega_t|^{-1}$ in the high-frequency regime. 
Moreover, we recover the expected scaling in $|\omega_t|^{-2}$ for the turbulent viscosity $|\Re_e(\widehat{\nu})_E|$ in the high-frequency regime (in addition to a linear reduction factor in an intermediate regime), which is always smaller than $|\Im_m(\widehat{\nu}_E)|$. 
The effective viscosity has approximately the same amplitude when it is calculated using (\ref{eq:closurexy}) or equation (\ref{eq:eddy}), so this cross-validates our computations for the turbulent viscosity. We have checked that quantitatively similar results are obtained by considering the other components 
$\langle u_x^2\rangle_V$ and $\langle u_y^2\rangle_V$ of the Reynolds stress tensor.
This agrees with \citet{penev2009direct}, who showed 
that the effects of convective turbulence on a large-scale oscillatory shear flow is well represented by an effective viscosity coefficient.

\section{Estimates for Tidal Synchronization}
\label{appendix:extrapolation}
We provide details here to compute the tidal dissipation timescales shown in Figure \ref{fig:eclipsing}. 
For a circular and aligned orbit, the turbulent dissipation is estimated from a spherical stellar model as \citep[e.g. see Equation (85) in][]{remus2012equilibrium}
\begin{equation}
    \mathcal{D}_\nu = 4\pi\frac{2088}{35} \frac{R^4}{GM}|\omega_t|\int_{\alpha_{R}}^1 x_R^8 \, \rho_*  \nu_E \,\mathrm{d} x_R,
\end{equation}
where $x_R=r/R$ is the normalized radius, $M$ is the stellar mass, $\alpha_{R}$ is the ratio of the radius of the base of the convective envelope to the stellar radius $R$, and $\rho_*$ is the density. 
To obtain leading-order estimates, we use the stellar models from EZ-Web (\url{http://www.astro.wisc.edu/~townsend/static.php?ref=ez-web}) for a 1 solar-mass star at 1 Gyr (assuming the metallicity $Z=0.02$). 
The modified tidal quality factor $Q'$ is related to $\mathcal{D}_\nu$ by $Q'=3/(2\mathcal{D}_\nu)$. 
The resulting timescale for tidal synchronization of the stellar spin is then
\begin{equation}
    \tau_\Omega = \frac{1}{3\pi  r_g^2}\left(\frac{M+M_2}{M_2}\right)^2\frac{P_{orb}^4}{P_{dyn}^2 \, P_s}\frac{1}{\mathcal{D}_\nu},
    \label{eq:timesync}
\end{equation}
where $M_2$ is the mass of the companion, $P_{orb}$ is the orbital period, $P_s$ is the stellar rotation period, $P_{dyn}=2\pi/\sqrt{GM/R^3}$ is the dynamical timescale, and $r_g^2\approx 0.1$ is the dimensionless squared radius of gyration. 

\begin{figure}
    \centering
    \begin{tabular}{c}
    \includegraphics[width=0.38\textwidth]{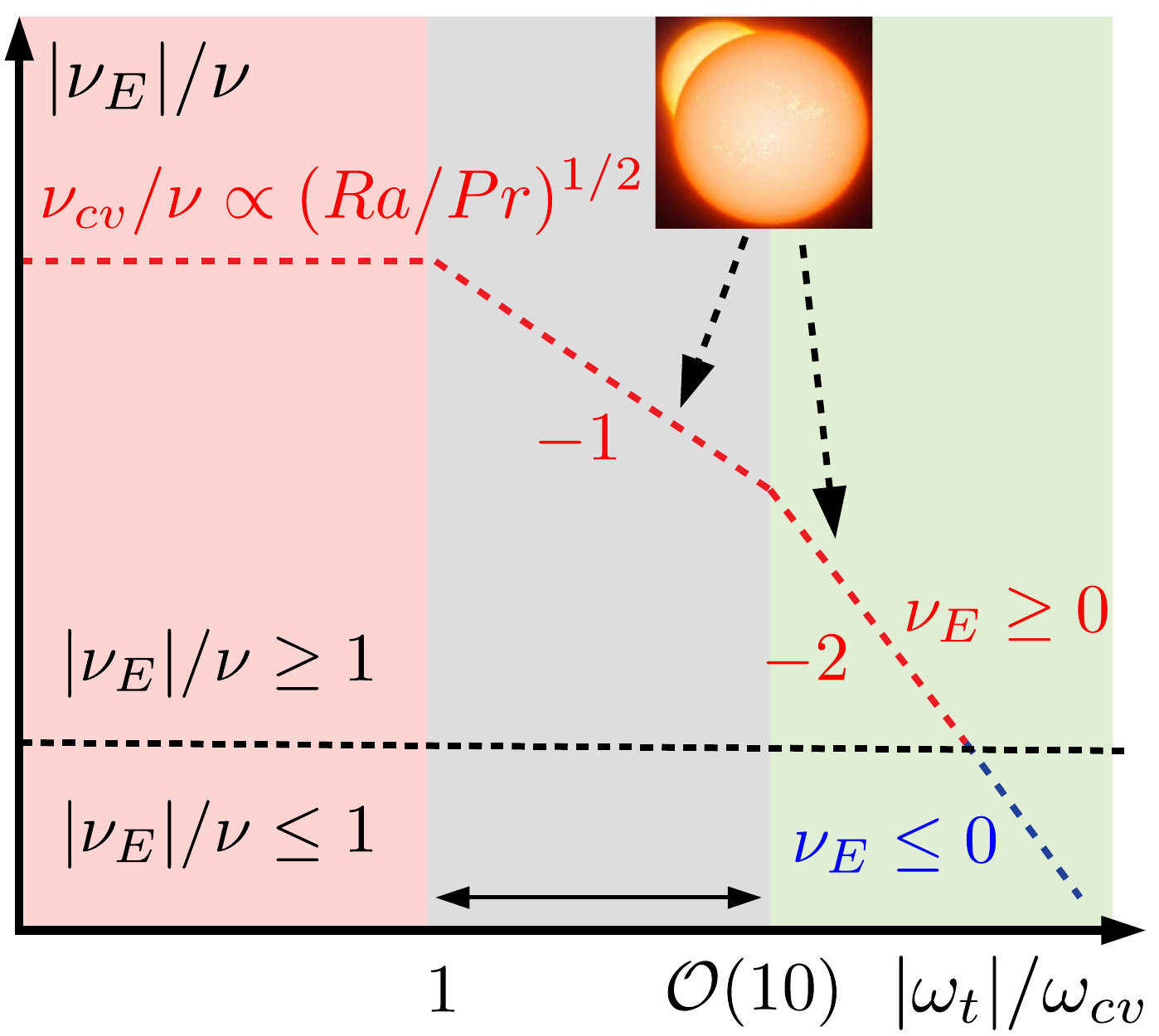} \\
    \end{tabular}
    \caption{Expected behavior of $|\nu_E|$, as a function of $|\omega_t|/ \omega_{cv}$, in turbulent stellar interiors. Laminar viscosity $\nu$ and turbulent one $\nu_{cv} \simeq u_{rms} \, l_E$ (MLT). Background colors refer to Figure \ref{fig:eclipsing}.}
    \label{fig:cartoon}
\end{figure}

Our simulations support the coexistence of the two frequency-reduction laws.
To evaluate equation (\ref{eq:timesync}),
we first estimate (from the stellar model) the convective velocity $u_{cv}(x_R)$, the mixing length $l_E(x_R)=2 \,H_p(x_R)$ with $H_p(x_R)$ the pressure scale height, and the convective frequency $\omega_{cv}(x_R)=u_{cv}(x_R)/l_E(x_R)$.
Then, we assume that the frequency reduction of the turbulent viscosity $\nu_E(x_R)$ obeys the continuous profile (Figure \ref{fig:cartoon})
\begin{equation}
 \nu_E =  u_{cv} \ell_E \begin{cases}
 1 \quad & (|\omega_t|/\omega_{cv}<1), \\
 {\omega_{cv}}/{|\omega_{t}|} \quad &(|\omega_t|/\omega_{cv} \in [1,5]), \\
 5\left({\omega_{cv}}/{|\omega_{t}|}\right)^2 \quad &(|\omega_t|/\omega_{cv}>5), \\
 \end{cases}
 \label{eq:continuousprofile}
\end{equation}
where the proportionality constant is arbitrary but is chosen here to be consistent with Figure \ref{fig:eddy}. 
The apparent discontinuity, reported in Figure \ref{fig:eddy}, apparently coincides with the rapid passage through zero of $\nu_E$ in the simulations. Since negative values of $\nu_E$ may not be relevant in reality in the frequency range $|\omega_t|/\omega_{cv} \leq 100$, we adopt here a continuous frequency-reduction profile. 

Taking a solar-mass binary with $P_\mathrm{orb}=10$ days and $P_s\sim$ 15 days (for example), we would obtain $\tau_\Omega \approx 1.14$ Gyr.
This timescale should be compared with $\tau_\Omega\approx 100$ Myr from neglecting the frequency reduction of $\nu_E$.
The resulting synchronization timescales for profile (\ref{eq:continuousprofile}) obtained using Equation (\ref{eq:timesync}) are superimposed in Figure \ref{fig:eclipsing}. 
Our extrapolation indicates that convective damping of the equilibrium tide can be important in driving spin synchronization in the sample presented in \citet{lurie2017tidal}. 

\bibliographystyle{aasjournal}
\bibliography{main}{}

\end{document}